\documentstyle[11pt]{article}

\def\bc{\begin{center}}
\def\ec{\end{center}}
\def\be{\begin{equation}}
\def\ee{\end{equation}}
\def\myappendix{\par
 \setcounter{section}{0}
 \setcounter{subsection}{0}
 \setcounter{equation}{0}
 \setcounter{table}{0}
 \def\appendixname{Appendix}
 \def\appesection{\setcounter{equation}{0}\section}
 \def\@thesection{\Alph{section}}
 \def\thesection{\appendixname\hskip 1.10ex\Alph{section}}
 \def\thesubsection{\@thesection.\arabic{subsection}}
 \def\theequation{\@thesection.\arabic{equation}}
 \def\thetable{\@thesection.\arabic{table}}}

\newcommand{\mfrac}[2]{\frac{\textstyle #1}{\textstyle #2}}

\newcommand{\alpi}{\left( \mfrac{\alpha_{s}}{\pi} \right)}
\newcommand{\alpicf}{\left( \mfrac{\alpha_{s}}{\pi} \right)\, \mfrac{C_{F}}{4}}
\newcommand{\aalpicf}{\left( \mfrac{\alpha_{s}(a^{-1})}{\pi} \right)\,
\mfrac{C_{F}}{4}}

\newcommand{\opv}{(\vec{v} \cdot \vec{D})}

\newcommand{\uno}{\mbox{\bf 1}}
\newcommand{\mkappa}{\mbox{\bf K}}

\newcommand{\beq}{\begin{equation}}
\newcommand{\eeq}{\end{equation}}
\newcommand{\beqn}{\begin{eqnarray}}
\newcommand{\eeqn}{\end{eqnarray}}

\def\vdir{v\kern-7.8pt\Big{/}}
\def\pdir{p\kern-7.8pt\Big{/}}
\newcommand{\diff}{\rm d}
\begin{document}
\titlepage
\bc
{\huge \bf Renormalization of the effective theory for heavy quarks at
                                                    small velocity}
\ec
\vskip 1.2cm
{\Large U.~AGLIETTI}\footnote{INFN, Sezione di Roma,
at present at Caltech, Pasadena, USA. E-mail: VAXROM::AGLIETTI.}
and
{\Large V.~GIM\'ENEZ}\footnote{On leave of absence from the Department of
Theoretical Physics and IFIC, Universitat de Val\`encia--CSIC, E-46100
Burjassot (Valencia),
Spain. E--mail: Decnet EVALVX::GIMENEZ.}\footnote{CEE Fellow.
Work partially supported by CICYT AEN 93-0234.}\\
{\it Dipartimento di Fisica ''G. Marconi'', Universit\`a degli studi di Roma
''La Sapienza'',\\Piazzale Aldo Moro 2, I-00185 Roma, Italy.}\\
\vskip 0.8cm
\bc
{\bf Abstract}
\ec
The slope of the
Isgur-Wise function at the normalization point, $\xi^{(1)}(1)$,
is one of the basic parameters for the extraction of the $CKM$ matrix element
$V_{cb}$ from exclusive semileptonic decay data.
A method for measuring
this parameter on the lattice is the effective theory for
heavy quarks at small velocity $v$.
This theory is a variant of the heavy quark effective theory
in which the motion of the quark
is treated as a perturbation.
In this work we study the lattice renormalization of
the slow heavy quark effective theory.
We show that the renormalization
of $\xi^{(1)}(1)$ is not affected by ultraviolet power divergences,
implying no need of difficult non-perturbative subtractions.
A lattice computation of $\xi^{(1)}(1)$ with this method is therefore
feasible in principle.
 The one-loop renormalization constants of the
effective theory for slow heavy quarks are computed to order
$v^2$ together with
the lattice-continuum
renormalization constant of $\xi^{(1)}(1)$ .

We demonstrate that the expansion in the heavy-quark
velocity reproduces correctly the infrared structure
of the original (non-expanded) theory to every order.
We compute also the one-loop renormalization constants of the
slow heavy quark effective theory to higher orders in $v^2$
and the lattice-continuum renormalization
constants of the higher derivatives of the $\xi$ function.
Unfortunately, the renormalization constants of the higher
derivatives are affected by ultraviolet power divergences,
implying the necessity of numerical non-perturbative subtractions.
The lattice computation of higher derivatives of the Isgur-Wise function
seems therefore problematic.
\vfill
\eject
\pagestyle{plain}
\bc
\section{Introduction} \label{intro}
\ec

The effective theory for heavy quarks $(HQET)$ \cite{eihil,geo}
(for a comprehensive review and references to the  original literature
see ref.\cite{Neubert})
allows a clean determination
of the Cabibbo-Kobayashi-Maskawa
matrix element $\mid V_{cb} \mid$ from the exclusive semileptonic
decays of $B$ mesons
\beq\label{eq:decay}
B ~ \rightarrow ~ D^{(*)}+l+\nu_l
\eeq
recently measured by the ARGUS \cite{argus} and CLEO \cite{cleo}
collaborations.

In the limit of infinite mass for the charm and the beauty quark,
\beq
m_c,~m_b\rightarrow\infty,
\eeq
the six form factors parametrizing the
hadronic matrix elements of the decays (\ref{eq:decay}) can all be
expressed in terms of a unique form factor, the Isgur-Wise function
$\xi(v\cdot v')$ \cite{isgw,isgwr,man},
\beqn\label{eq:iwf}
\langle D,v\mid J^{b \rightarrow c}_{\mu}(0)\mid B,v'\rangle  &=&
\sqrt{M_DM_B}~
(v_{\mu}+v'_{\mu})~\xi(v\cdot v')
\\
\langle D^*,v,\epsilon \mid J^{b \rightarrow c}_{\mu}(0)\mid B,v'\rangle
&=&
-\sqrt{M_DM_B}~[
i\epsilon_{\mu\nu\alpha\beta}\epsilon^{\nu}{v'}^{\alpha}v^{\beta}
+\epsilon_{\mu}(1+v\cdot v')-v_{\mu}v'\cdot \epsilon ]\xi(v\cdot v')
\label{eq:iwf2}
\eeqn
where $v'$ and $v$ denote respectively the $b$ and $c$ quark 4-velocities and
$J_{\mu}^{b\rightarrow c}(x)$ is the weak current describing the
transition of a beauty quark into a charm quark,
$J_{\mu}^{b\rightarrow c}(x)=\overline{c}(x)\gamma_{\mu}(1-\gamma_5)b(x)$.

\noindent
This function is normalized at zero recoil
\beq
\xi(v'\cdot v=1)~=~1,
\eeq
and $1/m$-corrections vanish in this kinematical point \cite{luke}.
A model independent analysis of the decays (\ref{eq:decay})
extrapolates  the experimental data up to the endpoint, where the
form factors are known by symmetry.
In order to eliminate the systematic errors introduced by the extrapolation,
it is essential to know also the derivative
of the Isgur-Wise function at the normalization point, $\xi^{(1)}(1)$.

So far, four methods have been proposed to compute on the lattice the slope
of the Isgur-Wise function. The first one was the estimation of the derivatives
of the Isgur-Wise function at the zero-recoil point from the lattice
determination of the Isgur-Wise function at discrete points in a region close
to the zero-recoil point \cite{altri}. This method presents some problems
related to the extrapolation to the zero-recoil point that
could lead to uncertainties in the determination of $\xi^{(1)}(1)$.
The authors of \cite{ugo2} have suggested a new method to compute directly
on the lattice the slope of the Isgur-Wise function which avoids any kind of
extrapolation. They proposed to study the first spatial moments of two- and
three-point meson correlators. Then the derivatives of the Isgur-Wise function
could be extracted by forming appropriate ratios of these correlators. The
UKQCD
Collaboration has recently carried out a exploratory study of the feasibility
of this method \cite{altri2}. The main conclusion is that
there are large finite-volume effects
in the lattice
evaluation of the moments of the correlation functions,
having a geometrical origin.
Therefore, by increasing the
length of the lattice in the spatial directions, these undesiderable volume
effects can be reduced. Unfortunately, they are large on currently available
lattices. Some approximations have been presented by this group in order to
control the volume effects and extract the slope on finite volumes.

Both computations described
above treat the heavy quark as an ordinary quark but with a
small hopping constant. The first calculation of the Isgur-Wise function
using the lattice $HQET$ has been
done by the authors of \cite{mandula2}. In that
work, the lattice propagator of the heavy quark with velocity $v$ is obtained
from a Wick rotated lagrangian \cite{mandu}.

The fourth method is based on
an expansion of the $HQET$ in the heavy-quark velocity around the static
theory (hereafter called 'effective theory for slow
heavy quarks', $SHQET$) \cite{ugo1}.
As in the method of spatial moments \cite{ugo2}, the derivatives of the
Isgur-Wise function at the zero-recoil point can be extracted directly
from ratios of two- and three-point correlation functions (see Section 2 for
details). The main point is that
there are no unexpected geometrical volume
effects
in the lattice
computation of these correlators because the static propagator
is local in space.
Moreover, the $SHQET$ circumvents the problem of the
euclidean  continuation of the Georgi theory for
heavy-quarks with non-vanishing velocity \cite{mandu,noi}. In fact, in order
to simulate heavy-quark with velocity $v$ on the lattice, the continuum
Minkowskian $HQET$ must be transformed
into a discretized euclidean field theory.
The analytical continuation is not simple because the energy spectrum
is unbounded from below \cite{mandu,noi,real}.
On the contrary,
expanding around small velocities, we are perturbing the static
theory whose energy
spectrum is bounded from below.
Roughly speaking, we may say that the heavy quark has a 'perturbative motion'
in the $SHQET$ produced by the 'velocity operator' $\opv$. This
theory has not been used yet in numerical lattice simulations.

In this paper, we analyse the lattice version of the $SHQET$.
One of our main results
is that the lattice renormalization
constant of $\xi^{(1)}(1)$ does not contain any ultraviolet
power divergence (i.e. proportional to $1/a^n$, where $a$ is the lattice
spacing).
The renormalization of $\xi^{(1)}(1)$ is affected only by logarithmic
divergences (of the form $\log a m $)
which can be subtracted with ordinary perturbative computations.
This implies that the lattice computation of $\xi^{(1)}(1)$
by simulations using the $SHQET$ is feasible in principle.
Ultraviolet power divergences are indeed a serious problem
for numerical simulations, because they cannot be subtracted
perturbatively with adequate accuracy \cite{bochi,guido}.
We also compute the one-loop lattice renormalization
constant of $\xi^{(1)}(1)$. The knowledge of this
renormalization constant is essential for converting the values
of $\xi^{(1)}(1)$ computed on the lattice to the values in the
original (high-energy) theory.
Moreover it is shown
that the infrared as well as the ultraviolet behaviour of the non-expanded
theory $(HQET)$ are reproduced by the $SHQET$ order by order in the velocity.
This is a non trivial check of the consistency of our approach.
We also give the lattice renormalization
constants of higher derivatives of the Isgur-Wise function,
$\xi^{(n)}(1)$, $n>1$.
Unfortunately, the lattice renormalization constants of higher derivatives
$\xi^{(n)}(1)$ for $n>1$ are affected by power divergences.
The computation of higher derivatives with $SHQET$ is therefore
more difficult than that one of $\xi^{(1)}(1)$.

This paper is organized as follows.
In section 2 we review the $SHQET$ in the
continuum \cite{ugo1};
the derivatives of the Isgur-Wise function are expressed as
ratios of three- and two-point correlation
functions.
Section 3 deals with the lattice regularization of the $SHQET$.
In section 4
we briefly review the matching theory of lattice operators
onto the continuum ones.
In section 5 we
renormalize the lattice $SHQET$ at order $\alpha_s$
and to all orders in the velocity.
In section 6 the renormalization of the heavy
quark current $J_{\mu}^{b\rightarrow c}$ is computed.
In section 7 we calculate the
lattice-continuum renormalization constants for the derivatives of the
Isgur-Wise function. Section 8
deals with the problem of power divergences.
Finally, in section 9 we present our conclusions.
There are also two appendices where the technique to compute lattice
integrals is described.
In appendix A the
analytical expressions and numerical values
of one-loop diagrams are
presented and in appendix B the method for subtracting infrared divergences is
briefly explained.

\section{The effective theory for slow heavy quarks}

In this section we review  the basic results and formulas of the
$SHQET$.
The Georgi lagrangian describing a heavy quark $Q$ with velocity
$v^{\mu}=(\sqrt{1+\vec{v}^2},\vec{v})$ in Minkowsky space \cite{geo}
\beq
{\cal L}(x)~=~Q^{\dagger}(x)i v\cdot D(x)Q(x)
\eeq
is decomposed as
\beq
{\cal L}(x)~=~{\cal L}_0(x)~+~{\cal L}_I(x)
\eeq
where
\beq
{\cal L}_0(x)~=~Q^{\dagger}(x)iD_0(x)Q(x)
\eeq
is the static unperturbed lagrangian and
\beq
{\cal L}_I(x)~=~Q^{\dagger}(x)i[D_0(v_0-1)-\vec{v}\cdot\vec{D}]Q(x)
\eeq
is a perturbation lagrangian giving rise to the motion of $Q$.

\noindent
{}From this splitting it is easy to derive the following expansion
of the propagator of $Q$ \cite{ugo1}
\beqn\nonumber
S(x,y;v)&=&-i~\Theta(t_x-t_y)~\Bigg(P(t_x,t_y)
+\int_{t_y}^{t_x}dt_{z}P(t_x,t_z)\vec{v}\cdot\vec{D}(\vec{x},t_z)
P(t_z,t_y)
\\ \nonumber
&+&\int_{t_y}^{t_x}dt_{z}\int_{t_y}^{t_z}dt_w
P(t_x,t_z)\vec{v}\cdot\vec{D}(\vec{x},t_z)
P(t_z,t_w)\vec{v}\cdot\vec{D}(\vec{x},t_w)P(t_w,t_y)
\\
& &~~~~~~~~~~~~~~~~~~~~~~~~~~~~-\frac{v^2}{2} P(t_x,t_y)+\ldots~\Bigg)~
\delta(\vec{x}-\vec{y})
\label{eq:soluz}
\eeqn
where $P(t_b,t_a)$ is a P-line in the time direction
joining the point $(\vec{x},t_a)$ with the point $(\vec{x},t_b)$
\beq
P(t_b,t_a)~=~P\exp\Bigg(ig\int_{t_a}^{t_b}A_0(\vec{x},s)ds\Bigg)
\eeq
$S(x,y;v)$ is expressed as a sum of static propagators
with an increasing number of local insertions of $\opv$ giving rise
to the 'perturbative' motion of $Q$.

The spin structure of $Q$ is taken into account multiplying
$S(x,y;v)$ by $(1+\vdir)/2$
\beqn \label{propa}
\nonumber
H(v)&=&\frac{ 1+\vdir }{2}~S(v)
\\ \nonumber
&=&\frac{1+\gamma_0}{2}S^{(0)} +
\Bigg(\frac{1+\gamma_0}{2}S^{(1)}-\frac{\gamma_3}{2}S^{(0)}\Bigg)\, v_{3}
\\
&+&\Bigg( \frac{1+\gamma_0}{2}S^{(2)}-\frac{\gamma_3}{2}S^{(1)}
+\frac{\gamma_0}{4}S^{(0)} \Bigg)\, v_{3}^{2}\,+\, O(v_{3}^3)
\eeqn
where we have taken the heavy-quark moving along the $z$-axis.
\noindent
Inserting the propagator $H(v)$ in the Green's functions describing
the dynamics of heavy flavored hadrons, we have the following
expansion in powers of $v_{3}$
\beq \label{exp}
G(v)~=~G^{(0)}~+~G^{(1)}~v_{3}~+~G^{(2)}~v_{3}^2+\cdots
\eeq

Consider now the following three- and two-point correlation functions
\beqn
C_3(t,t')&=&\int d^3xd^3x'
\langle 0\mid T~[O_D^{\dagger}(x'),J^{b\rightarrow c}_{\mu}(x),
O_B(0)] \mid 0\rangle
\label{eq:3pt}
\\
C_B(t)&=&\int d^3x
\langle 0\mid T~[O_B^{\dagger}(x),O_B(0)] \mid 0\rangle
\label{eq:2B}
\\
C_D(t'-t)&=&\int d^3x'
\langle 0\mid T~[O_D^{\dagger}(x'),O_D(x)] \mid 0\rangle
\label{eq:2D}
\eeqn
where
$O_{H}(x)$ is an interpolating field for the $H$ meson.
The simplest choice (which we adopt in the following) is:
$O_{H}(x)=\overline{Q}(x)i\gamma_5 q(x)$, where $Q(x)=b(x),~c(x)$
 for $B$ and $D$ mesons respectively,
and $q(x)$ is a light quark field.
\noindent
For large euclidean times, $t\rightarrow\infty$, $t'-t\rightarrow\infty$,
the matrix element (\ref{eq:iwf}) is given by
\beq
     \langle D,v\mid J^{b\rightarrow c}_{\mu}(0)\mid B,v'\rangle~
=~\sqrt{Z_B~Z_D}~
\frac{C_3(t,t')}{C_B(t)~C_D(t'-t)}
\label{eq:rat}\eeq
where $Z_{B}$ and $Z_{D}$ are the renormalization constants of the
operators $O_{B}(x)$ and $O_D(x)$, given by
\beqn\nonumber
\sqrt{Z_{B}}&=&\langle B,v'\mid O_{B}(0)\mid 0\rangle
\\
\sqrt{Z_{D}}&=&\langle D,v\mid O_{D}(0)\mid 0\rangle
\label{eq:rc}\eeqn
Since both the wave functions and the interpolating fields
in eqs.(\ref{eq:rc}) are pseudoscalars,
the matrix elements do not depend on the velocity
$v(v')$, unless we deal with smeared currents \cite{ugo2}.

Inserting the propagator (\ref{propa}) for the
$c$ quark in $C_{3}(t,t')$ and $C_{D}(t'-t)$ we derive expansions of the
form (\ref{exp}).
Inserting them into eq.(\ref{eq:rat}), we get the following expression
for the derivatives of the Isgur-Wise function with respect to
$v_{4}$ at the zero recoil point $v_{4}=1$
\begin{eqnarray} \label{ratios1}
\left[\,
\mfrac{C_3^{(2)}}{C_3^{(0)}}\, -\, \mfrac{C_D^{(2)}}{C_D^{(0)}}\, \right]
&=& \mfrac{1}{2}(\xi^{(1)}(1)+\mfrac{1}{2})\\
\label{ratios2}
\left[\,
\mfrac{C_3^{(4)}}{C_3^{(0)}}\, -\, \mfrac{C_D^{(4)}}{C_D^{(0)}}\,
+\, \left(\mfrac{C_D^{(2)}}{C_D^{(0)}}\right)^{2}\, -\,
\mfrac{C_D^{(2)}}{C_D^{(0)}}~\mfrac{C_3^{(2)}}{C_3^{(0)}}\, \right]
&=&
\mfrac{1}{4}(\xi^{(2)}(1)-\mfrac{1}{2})
\end{eqnarray}
where we have used the identity
\be
\mfrac{C_3^{(0)}}{ C_B~C_D^{(0)}}\, =\, \sqrt{ \frac{2M_D~2M_B}{Z_D~Z_B} }
\ee
Higher derivatives can be computed similarly.

\section{Lattice regularization}

We consider the discretization of the effective theory for heavy
quarks proposed in ref.\cite{mandu}, forward in time and
symmetric in space. For a motion of $Q$ with velocity
along the $z$-axis $v^{\mu}=(0,0,v_{3},\sqrt{1+v_{3}^2})$,
the action $iS$ is given by
\beqn
iS &=&  -\sum_x ~
v_4\psi^{\dagger}(x)~[\psi(x)-U_4^{\dagger}(x)\psi(x-\vec{4})]+~
\nonumber\\
 & &~~~~~-i\frac{v_3}{2}\psi^{\dagger}(x)~[U_3(x+\vec{3})\psi(x
+\vec{3})
                                 -U_3^{\dagger}(x)\psi(x-\vec{3}) ]
\label{eq:act}\eeqn
where $\vec{\mu}$ is a unit vector in the direction $\mu$, and
$U_{\mu}(x)$
are the links related to the gauge field by
$U_{\mu}(x)~=~\exp[-igA_{\mu}(x-\vec{\mu}/2)]$.

The Feynman rules are those of the static theory plus
additional interactions generating the motion of $Q$.
Assuming a convention for the Fourier transform according to which
$\psi(x)\sim \exp(ik\cdot x)$, we have
\beqn\label{propa0}
  iS^{(0)}(k)&=& \frac{1}{1-e^{-ik_4}+\epsilon}
\\
V^{(0)}_{\mu}&=& i\,g\,\delta_{\mu 4}\, t_a~e^{-i(k_4+{k'}_4)/2}
\\
V^{(0)~tad}_{\mu\nu}&=&-\frac{g^2}{2}\,\delta_{\mu 4}\delta_{\nu 4}\,
                                    t_at_b~e^{-ik_4}
\eeqn
The linear interactions in $v_{3}$ are given by
\beqn
\label{eq:vthree1}
   V^{(1)}&=& -v_{3}\,\sin k_3\\ \label{eq:act1}
V^{(1)}_{\mu}&=&g\, v_{3}\, \delta_{\mu 3}\,t_a\, \cos(k_3+{k'}_3)/2\\
\label{eq:vthree2}
V_{\mu\nu}^{(1)~tad}&=&\frac{g^2\, v_{3}}{2}\,
\delta_{\mu 3}\delta_{\nu 3}\,t_at_b\,\sin k_3
\eeqn
and the linear interactions in $(v_{4}-1)$ are given by
\beqn
\label{eq:vfour1}
V^{(2)}&=&-(v_{4}-1)(1-e^{-ik_4})\\
\label{eq:act2}
V_{\mu}^{(2)}  &=& i\, g\, (v_{4}-1)\, \delta_{\mu 4}\,t_a\,
e^{-i(k_4+{k'}_4)/2}\\
\label{eq:vfour2}
V_{\mu\nu}^{(2)~tad}&=& -\frac{g^2}{2}\, (v_{4}-1)\,
\delta_{\mu 4}\delta_{\nu 4}\, t_a t_b\, e^{-ik_4}
\eeqn
where $k$ and $k'$ denote respectively the momenta of the incoming
and outgoing heavy quark, and $V_{\mu}$ is the interaction
vertex of the heavy quark with a gluon provided with a
polarization along the $\mu$ axis.
$V^{tad}_{\mu\nu}$ are the vertices for the emission of
two gluons, for the case of the tadpole graph ($k=k'$).
We notice that the vertices labelled $V^{(2)}$
contain second and higher
orders in the velocity $v_{3}$. It is convenient to
keep them unexpanded.
Finally, note that the conventions for the sign of the Fourier transform
and of the velocity are not independent, if one wishes to intend
$k$ as the residual momentum of the heavy quark.

\section{Renormalization of lattice operators}

Since the lattice effective theory and the continuum one are two
different versions of the same physical theory, the matrix elements
computed in both theories must coincide. This is a non-trivial
condition to impose (matching condition).
We match amplitudes of the bare lattice theory onto the corresponding
ones of the continuum theory renormalized in some chosen scheme
(such as for example $\overline{MS}$).
If the lattice lagrangian and the continuum one have at the
beginning the same parameters
(masses, couplings, etc.), matching is accomplished adding
appropriate counterterms to the lattice lagrangian.
If we are interested also in the matrix elements of composite operators,
an analogous matching has to be performed: appropriate counterterms
have to be added to the lattice composite operators.
Because of mixing, to a renormalized operator in the
continuum it corresponds in general
a linear combination of lattice bare operators.

The technique for obtaining the lattice counterpart of a continuum operator
is standard \cite{nonper1,opelat}.
In lattice regularization the inverse of the lattice spacing
$1/a$ acts as an ultraviolet cut-off, and bare lattice amplitudes depend
explicitely on $a$. Continuum amplitudes depend instead on a renormalization
point $\mu$.
To avoid large logarithms in the matching constants
($\log a\mu >>1$), let us first match
the amplitudes by taking $\mu=a^{-1}$.
At this stage we are therefore dealing with
the finite discrepancies
coming from the use of different regulators.
The relation between continuum and lattice
operators at one-loop level for $\mu=a^{-1}$ is given by
\be \label{xx1}
O_{i}^{Cont}(\mu=a^{-1})\, =\, \sum_{j}\, [\, \delta_{i j}\, +\,
\left(\mfrac{\alpha_{s}(a^{-1})}{\pi}\right)\, \delta Z_{i j}\, ]\,
O_{j}^{Latt}(a)
\ee
where $O_{i}^{Cont}$ are the operators in the continuum we are interested in,
$O_{j}^{Latt}$ are the lattice ones and $\delta Z_{i j}$ are
finite renormalization (or mixing) constants.
The sum extend over
all the operators
that can mix with $O_{i}^{Latt}$
as a consequence of the symmetry breaking induced by
the continuum and the lattice regularization procedure.
By sandwiching the operators between arbitrary external states of momenta
$p$, we derive the following relation involving
Green's functions of the bare lattice operators or the (renormalized)
continuum ones
\be \label{xx2}
<\, O_{i}^{Cont,Latt} \,>\, =\, \sum_{j}\, [\, \delta_{i j}\, +\,
\left(\mfrac{\alpha_{s}(a^{-1})}{\pi}\right)\, C^{Cont,Latt}_{i j}(p)\, ]\,
<\, O_{j}\, >^{(0)}
\ee
where the superscript $(0)$ denotes tree level matrix elements.
Demanding compatibility between (\ref{xx1}) and (\ref{xx2}), we derive
\be
\delta Z_{i j}\, =\, \lim_{a\rightarrow 0} \left[\,
C^{Cont}_{i j}(p)\, -\, C^{Latt}_{i j}(p)\, \right]
\ee
The mixing coefficients $\delta Z_{ij}$ are
independent of both the external states and the momentum configuration used to
calculate matrix elements of $O_{i}$.

We consider now the matching in the more general case $\mu a\neq 1$.
Since we already matched the amplitudes at $\mu a =1$,
we need only to evolve the mixing coefficients $\delta Z_{ij}$
from $\mu=1/a$ to a generic renormalization point with $RG$ techniques.
At one loop-level, the $\delta Z_{ij}$'s do
depend on the renormalization scheme.
In order to obtain a renormalization scheme independent
matching condition, the
two-loop anomalous dimension contribution must be taken into
account in the diagonal terms \cite{twoloop,twoloop2}
\begin{eqnarray} \label{xx3}
O_{i}^{Cont}(\mu) &=&
\left(\mfrac{\alpha_{s}(a^{-1})}{\alpha_{s}(\mu)}
\right)^{\gamma_{1}/\beta_{1}}\, [\, 1\, +\,
\left(\mfrac{\alpha_{s}(a^{-1})}{\pi}\, -\,
\mfrac{\alpha_{s}(\mu)}{\pi}\right)\, R_{O_{i}}\, ]\nonumber\\
&\times& \sum_{j}\, [\, \delta_{i j}\, +\,
\left(\mfrac{\alpha_{s}(a^{-1})}{\pi}\right)\, \delta Z_{i j}\, ]\, O_{j}
^{Latt}(a)
\end{eqnarray}
where
\be
R_{O_{i}}\, =\, \mfrac{1}{\beta^{2}_{1}}\, \left[\,
\gamma_{2}\, \beta_{1}\, -\, \gamma_{1}\, \beta_{2}\, \right]
\ee
with $\gamma_{n}$ the n-loop anomalous dimension of the operator $O_{i}$
defined by
\begin{eqnarray}
O_{i}^{R} &=& Z_{O}\, O_{i}^{B}\nonumber\\
\gamma &=& - \mu\, \mfrac{{\rm d}}{{\rm d} \mu}\, \log Z_{O}\, =\,
 \gamma_{1}\, \alpi\, +\, \gamma_{2}\, \alpi^{2}\, +\, \cdots
\end{eqnarray}
and with $\beta_1$ and $\beta_2$ the one-loop and two-loop
coefficients of the $\beta$-function respectively
\beq
\beta(\alpha_s)~=~\beta_1\frac{\alpha_s}{\pi}+
          \beta_2\left(\frac{\alpha_s}{\pi}\right)^2+\ldots
\eeq
We have
\beqn
\beta_{1} &=& -\mfrac{11}{2} + \mfrac{1}{3} n_{F}\nonumber\\
\beta_{2} &=& -\mfrac{51}{4} + \mfrac{19}{12} n_{F}
\eeqn
The expression for the running coupling constant is given by
\beq
\alpha_{s}(\mu) = \mfrac{2 \pi}{- \beta_{1}\, \log (\mu^{2}/\Lambda^{2})}\,
\left[\, 1\, +\, \mfrac{2 \beta_{2}\, \log\,
\log(\mu^{2}/\Lambda^{2})}{\beta_{1}^{2}\, \log(\mu^{2}/\Lambda^{2})}\, \right]
\eeq
where $n_{F}$ is the number of active
quark flavors and we can take
$\Lambda= 200$ MeV in the $\overline{MS}$ scheme.

It is expected that for the
values of $a^{-1}$ currently used in lattice simulations the matching
should depend only weakly on the continuum regularization. This is
why physicists usually compute matching constants without including the
two-loop anomalous dimension.

Finally, let us briefly expose the problem of the power divergences
in lattice computations.
Since $QCD$ is asymptotically free,
the matching constants $\delta Z_{i j}$ can in principle be computed with
$RG$-improved
perturbation theory in the limit $a\rightarrow 0$
(in practise one requires $a\Lambda\ll 1$).
Unfortunately, the mixing coefficients contain in some cases
inverse powers of $a$,
 which divergence as $a$ goes to zero.
Then, in computations of matrix elements of the continuum operator,
the leading term is this
mixing term of $O(1/a^{n})$ which is a lattice artifact
generated by the regularization procedure and thus must be subtracted.
In other words, in order
to obtain finite Green functions of composite operators
on the lattice, we must subtract power divergences in $a^{-1}$ from the
Monte Carlo data. It has been argued that it can be done in perturbation
theory. However, as pointed out in refs.\cite{bochi,guido,beneke,bigi},
it is not clear that the coefficients of power divergences
can be calculated to sufficient accuracy
in perturbation theory. In general, very difficult non-perturbative
subtractions for lattice Green's functions are required.

\section{Renormalization of the lattice $SHQET$}

In this section we discuss the renormalization of the $SHQET$ given by the
lagrangian (\ref{eq:act}), i.e.~the determination of the counterterms which
have to be introduced to match amplitudes of the lattice $SHQET$ onto
those of the continuum $HQET$.

To obtain the renormalized operator $(\vec{v} \cdot \vec{D})$ we compute
the one-loop heavy quark self-energy
with insertions of $(\vec{v} \cdot \vec{D})$
using the lattice Feynman rules of section 3.
This is equivalent to calculate the one-loop heavy quark self-energy up to a
given order in the velocity $v_{3}$.

After that, we match the heavy quark propagator of the lattice $SHQET$
onto the continuum $HQET$ propagator,
expanded in $v_{3}$.

For calculational
convenience, we will take equal incoming and outgoing momenta
and adopt the Feynman gauge for the gluon propagator. The
infrared divergences which appear at
zero external momenta are regulated giving
the gluon a fictitious mass $\lambda$.
No problem arises with non-abelian gauge symmetry because
all the amplitudes are $QED$-like.
Other choices are possible for the infrared regulator, such as for example
to take virtual external states \cite{twoloop2}.
However, by using a
non-vanishing gluon mass, we achieve a great simplification in computing the
lattice loop integrals.
Indeed, we can safely Taylor expand the
corresponding diagrams about zero external
momenta up to order $O(a)$ to determine all nonvanishing terms as $a$ goes
to zero. Upon doing this, we will subtract the infrared (logarithmic)
divergences from the integrals with the technique explained in Appendix B.

The computation of the diagrams will be done with two different
methods for dealing with the non-covariant poles
coming from static lines.
The first method is based on partial integration with
respect to $k_{4}$ (the fourth-component of the euclidean loop momentum)
in order to eliminate the non-covariant poles.
The integrand is reduced to a covariant form and can be computed
with usual techniques.
The second method is to integrate analitically
over $k_{4}$ using the $\epsilon$-prescription of the
static heavy-quark propagator and the Cauchy's theorem.
This latter technique involves less algebra, but leads to
non-covariant 3-dimensional integrals.
The comparison of the results obtained with the two methods provides us with a
check not only of our analytical computation but also of our numerical
calculations.

\subsection{Heavy Quark Self-Energy up to $O(v_{3}^{2})$}

To illustrate the method of partial integration, in this section we briefly
describe the computation of the diagrams that determine the heavy-quark
self-energy up to $O(v_{3}^{2})$, which are depicted in Fig.1.
We do not consider the insertion of $v_{4}$-vertices (i.e.~those in
eqs.(\ref{eq:vfour1}) to (\ref{eq:vfour2})) because their contribution
can be shown to be trivial.
We will treat this subject in detail in the next section.

We start by computing the diagrams with one insertion of the operator $\opv$
in Fig.1.
Diagrams A.1 and A.2 vanish in the Feynman gauge. This happens
because the gluon is emitted by the operator $\opv$ with a polarization
along the $z$ axis, while it is absorbed by the static vertex with
a polarization along the time axis.
The (amputated) amplitude of diagram $A.3$ is given by
\beq
A_3(p)~=~g^2\, C_F\, v_{3}\,e^{-ip_4}\int_{-\pi}^{+\pi}\frac{d^4k}{(2\pi)^4}
\frac{ \sin k_3\, e^{-ik_4} }{ (1-e^{ik_4}+i\epsilon)^2 }~
\frac{1}{2\Delta_1(k-p)}
\eeq
where $p$ is the external momentum, $C_F=\sum t_at_a=(N^2-1)/2N$ for
an $SU(N)$ gauge theory, and
$\Delta_1(l)=\sum_{\mu=1}^4 1-\cos l_{\mu}+(a\lambda)^2/2$.

\noindent
Now, $A_{3}(p)$ vanishes at zero external momentum $p=0$,
\beq\label{eq:a3zero}
A_3(p=0)~=~0,
\eeq
because the integrand is odd in $k_3$.

\noindent
First derivatives of $A_{3}(p)$ with respect to the external momentum
contain logarithmic ultraviolet divergences.
The only non-vanishing derivative is that one with respect to $p_3$.
With a partial integration with respect to $k_4$
of the factor
\beq
\frac{ e^{-ik_4} }{(1-e^{-ik_4}+\epsilon)^2}
\eeq
we reduce the integral to the following form
\beq
\left(\frac{\partial A_3}{\partial p_3}\right)_0~
=~\frac{g^2C_F}{16\pi^2}\, v_{3}\, \frac{1}{6\pi^2}
\int_{-\pi}^{+\pi} \diff^4k~
\frac{\eta(\vec{k})~(1+\cos k_4) }{\Delta_1(k)^3}
\eeq
where $\eta(\vec{k})=\sum_{i=1}^{3}\sin^2 k_i$ and
a symmetrization over the spatial momenta has been done.
The infrared singularity of the integral is isolated
with the technique introduced in ref.\cite{twoloop2}
and described in detail in ref.\cite{real}. The result can be written as
\beq
\left(\frac{\partial A_3}{\partial p_3}\right)_0~
=~ \left(\mfrac{\alpha_{s}}{\pi}\right)\, \mfrac{C_F}{4}\, v_{3}\,
[~-2\log(a\lambda)^2+a_3~]
\eeq
where the subleading (finite) term $a_3$ is a constant evaluated numerically,
$a_3= 0.448$.

The tadpole graph $A.4$ is given by
\beq
A_4(p)~=~\frac{g^2\, C_F\, v_{3}}{4}\sin p_3\int_{-\pi}^{+\pi}
\frac{\diff ^4k}{(2\pi)^4}\frac{1}{\Delta_1(k)}
\eeq
As in the case of diagram $A.3$, $A_4(p)$ vanishes at zero external
momentum,
\beq\label{eq:a4zero}
A_4(p=0)=0.
\eeq
The first derivative with respect to $p_3$ is finite (i.e. does not
contain logarithmic divergences) and reads
\beq
\left(\frac{\partial A_4}{\partial p_3}\right)_0~
=~ \left(\mfrac{\alpha_{s}}{\pi}\right)\, \mfrac{C_F}{4}\,v_{3}~a_4
\eeq
where $a_4$ is a numerical constant, $a_4=12.23$.

Let us consider now the renormalization of a double insertion of $\opv$ at
zero momentum.
We study the Green function
\beq\label{eq:gprima}
G(z,w)~=~
\int d^4x d^4y~\langle 0\mid T[Q(z)\,Q^{\dagger}(x)\opv(x)Q(x)\,
Q^{\dagger}(y)\opv(y)Q(y)\,Q^{\dagger}(w)]\mid 0\rangle
\eeq
The diagrams involved are drawn in Fig.2. Diagrams $B.1$ and $B.2$ vanish in
the Feynman gauge.

\noindent
The amplitude of diagram $B.3$ can be written as
\beq
B_{3}(p)~=~-\frac{g^2\, C_F\, v_{3}^2}{6} e^{-ip_4}\int_{-\pi}^{+\pi}
\frac{\diff^4k}{(2\pi)^4}
\frac{ \eta(\vec{k})~e^{-ik_4} }{ (1-e^{-ik_4}+\epsilon)^3~\Delta_1(p-k) }
\eeq
The amplitude at zero external momentum is given by
\beq
B_3(0)~=~\frac{g^2\, C_F}{16\pi^2}\, v_{3}^2
\frac{-1}{6\pi^2}\int_{-\pi}^{+\pi}\diff ^4k~
\frac{ \eta(\vec{k})~e^{-ik_4} }{ (1-e^{-ik_4}+\epsilon)^3~\Delta_1(k) }
\eeq
It is convenient to reduce the integrand to a covariant form, by
eliminating the triple pole coming from the static line.
Let us describe in detail the transformation of this integral,
which will illustrate the technique to deal with poles of odd order.

We perform first a partial integration with respect to $k_4$
analogous to that one of $\partial A_3/\partial p_3$.
This transformation brings the integral into the form
\beq
B_{3}(0)~=~\frac{g^2\, C_F}{16\pi^2}\, v_{3}^2\frac{-1}{24\pi^2}
\int_{-\pi}^{+\pi}\diff ^4k
\frac{ \eta(\vec{k})~(e^{ik_4}+1) }{ (1-e^{-ik_4}+\epsilon) }
\frac{1}{ \Delta_1(k)^2 }
\eeq
The simple non-covariant pole is treated by writing \cite{eihil2}
\beq
\frac{1}{\Delta_1(k)^2}~=~\left(\frac{1}{\Delta_1(k)^2}
-\frac{1}{\Delta_1(0,\vec{k})^2}\right)~
+\frac{1}{\Delta_1(0,\vec{k})^2}
\eeq
In the integral containing the difference of gluon propagators,
\beq
I~=~\int_{-\pi}^{+\pi}\diff^4 k~
\frac{ 1+e^{ik_4} }{ 1-e^{-ik_4}+\epsilon }~
\eta(\vec{k})~
\left(\frac{1}{\Delta_1(k)^2}
-\frac{1}{\Delta_1(0,\vec{k})^2}\right),
\eeq
one can set $\epsilon=0$. Since the gluon propagator is even with respect
to $k_4$, one can symmetrize the factor
\beq
\frac{ 1+e^{ik_4} }{ 1-e^{-ik_4} }~\rightarrow~1+\cos k_4
\eeq
The integral $I$ therefore reads
\beq
I~=~\int_{-\pi}^{+\pi}\diff ^4 k~
\frac{ (1+\cos k_4)~\eta(\vec{k}) }{ \Delta_1(k) }~
-2\pi \int_{-\pi}^{+\pi}\diff ^3 k~\frac{\eta(\vec{k})}{ \Delta_1(0,\vec{k}) }
\eeq
In the remaining integral
\beq
J~=~\int_{-\pi}^{+\pi}\diff k_4\frac{ 1+e^{ik_4} }{ 1-e^{-ik_4}+\epsilon }~
\int_{-\pi}^{+\pi}\diff^3 k \frac{ \eta(\vec{k}) }{ \Delta_1(0,\vec{k})^2 },
\eeq
we perform  the contour integration over $k_4$ analytically by setting
$z=\exp(ik_4)$.
$B_3(0)$ is finally expressed as a sum of a 4-dimensional integral
and a 3-dimensional one
\beq\label{eq:a70}
B_3(0)~=~\frac{g^2C_F}{16\pi^2}v_{3}^2 \Bigg(-\frac{1}{24\pi^2}
\int d^4k\frac{ (1+\cos k_4)~\eta(\vec{k}) }{\Delta_1(k)^2}
-\frac{1}{12\pi}\int d^3 k
\frac{\eta(\vec{k})}{ \Delta_1(0,\vec{k})^2}\Bigg)
\eeq
The integrals in eq.(\ref{eq:a70})
are infrared finite and are easily computed numerically
\beq
B_{3}(0)~=~\alpicf\,v_{3}^2~b_{30}
\eeq
where $b_{30}=-5.044$.

\noindent
The first derivative of $B_3(p)$ with respect to $p_4$ is
logarithmically divergent, and is given by
\beq
\left(\frac{ \partial B_{3} }{ \partial p_4 }\right)_0~=~-iA_3(0)-
\frac{g^2\, C_F\, v_{3}^2}{6}\int\frac{ d^4k }{ (2\pi)^4 }
\frac{ \sin k_4~e^{-ik_4} }{ (1-e^{-ik_4}+\epsilon)^3 }
\frac{ \eta(\vec{k}) }{ \Delta_1(k)^2 }
\eeq
Using the same tricks as for $B_3(0)$, this integral is transformed into
\beqn\nonumber
\left(\frac{ \partial B_3}{\partial p_4}\right)_0&=&
-iB_3(0)+\frac{g^2C_F}{16\pi^2} v_{3}^2
\Bigg(~\frac{i}{12\pi^2}\int d^4 k
\frac{ \eta(\vec{k}) }{ \Delta_1(k)^3 }[1+2\cos k_4 +\cos 2k_4]
\\ \label{eq:der7}
&+&\frac{i}{12\pi^2}\int d^4 k \frac{ \eta(\vec{k}) }{ \Delta_1(k)^2 }
        [1/2+\cos k_4]
+\frac{i}{12\pi}\int d^3k \frac{ \eta(\vec{k}) }{ \Delta_1(0,\vec{k})^2 }~
\Bigg)
\eeqn
The logarithmic divergence of the amplitude
(the $\log(a\lambda)$ term)
is entirely contained in the first integral. The computation yields
\beq
\left(\frac{ \partial B_3 }{ \partial p_4 }\right)_0~=~
i\, \alpicf\, v_{3}^2~[~-b_{30}-2\log(a\lambda)^2+b_{31}~]
\eeq
where $b_{31}= 4.988$.

Finally, the amplitude of diagram $B_{4}$ is
\beq
B_4(p)~=~\frac{g^2 C_F\, v_{3}^2 }{ 12 }
\int_{-\pi}^{+\pi}\frac{\diff^4 k}{ (2\pi)^4 }~\frac{ 3+
\sigma(\vec{k}) }{ 1-e^{-ik_4}+\epsilon }~
\frac{1}{ \Delta_1(k-p) }
\eeq
where $\sigma(\vec{k})=\sum_{i=1}^3\cos k_i$.

\noindent
The computation of $B_4(p)$ is analogous to that of $B_3(p)$.
We have
\beq\label{eq:a8zero}
B_4(0)~=~\frac{g^2C_F}{16\pi^2}v^2 \Bigg( \frac{1}{24\pi^2}
\int \diff^4 k~\frac{ 3+\sigma(\vec{k}) }{ \Delta_1(k) }~
+~\frac{1}{12\pi}\int d^3 k~
\frac{ 3+\sigma(\vec{k}) }{ \Delta_1(0,\vec{k}) }  \Bigg)
\eeq

\noindent
Upon a numerical computation we find
\beq
B_4(0)~=~\alpicf\,v_{3}^2~b_{40}
\eeq
where $b_{40}=20.566$.

\noindent
The derivative with respect to $p_4$ reads
\beq
\left(\frac{ \partial B_{4} }{ \partial p_4 }\right)_0=
\frac{g^2C_F}{16\pi^2}\,v_{3}^2
\frac{-i}{24\pi^2}\int_{-\pi}^{+\pi} \diff^4 k
\frac{ [~3+\sigma(\vec{k})~][~1+\cos k_4]~ }{ \Delta_1(k)^2 }
\eeq
and the corresponding numerical computation yields
\beq
\left(\frac{ \partial B_4 }{ \partial p_4 }\right)_0~
=~i\,\alpicf\,v_{3}^2\, [~2\log(a\lambda)^2+b_{41}~]
\eeq
where $b_{41}=-2.485$.

Putting all contributions together, we can write the heavy-quark self-energy
up to order $O(v_{3}^{2})$ as
\begin{eqnarray} \label{oldse}
\Sigma(p,v)&=& \mfrac{1}{a}\, \alpicf\,
\left[\, \Sigma^{(0)}_{0}\, +\, v^{2}_{3}\, \Sigma^{(2)}_{0}\, \right]\, v_{4}
\nonumber\\
&+&\, \alpicf\, \left[\,
\Sigma^{(0)}_{40}\, +\, v_{3}^{2}\, \Sigma^{(2)}_{40}\, -\, 4\,
\log(a \lambda)\, \right]\, (i\, p_{4} v_{4})\nonumber\\
&+&\, \alpicf\, \left[\,
\Sigma^{(1)}_{30}\, -\, 4\, \log(a \lambda)\, \right]\,
(p_{3} v_{3})\nonumber\\
&+& O(\alpha_{s}\, a, v^{3}_{3})
\end{eqnarray}
where $\Sigma^{(0)}_{0}$ and $\Sigma^{(0)}_{40}$ are the mass and wave function
renormalization of a static heavy quark \cite{eihil2}. Their numerical values
are tabulated in Table A.1. On the other hand, the numerical values of the
new constants $\Sigma^{(1,2)}_{0,30,40}$ are
\begin{eqnarray} \label{oldsig}
\Sigma^{(2)}_{0} &=& b_{30}\, +\, b_{40}\, =\, 15.52\nonumber\\
\Sigma^{(2)}_{40} &=& b_{31}\, -\, b_{30}\, +\, b_{41}\, =\, 7.55\nonumber\\
\Sigma^{(1)}_{30} &=& a_{3}\, +\, a_{4}\, =\, 12.68
\end{eqnarray}
In the next section we compare (\ref{oldsig}) with the self-energy calculated
by using a different integration technique.

\subsection{Heavy Quark Self-Energy beyond $O(v_{3}^{2})$}

Here we compute the heavy-quark self-energy at one-loop
in the coupling constant $\alpha_{s}$ but to all orders in the velocity
$v_{3}$. We will demonstrate that the discretized lagrangian (\ref{eq:act})
reproduces
the correct infrared behaviour of the HQET to all orders in the velocity, as it
should be. The calculation will be performed utilizing a
different method from the one used in the previous section. This is useful
to check both our analytical and numerical results.

We start by noting that at one-loop the diagrams that contribute to the
self-energy of the heavy quark at order $O(v_{3}^{m})$ are those depicted in
Fig.3. In fact, these diagrams represent the only two ways of inserting $m$
$v_{3}$-vertices (i.e.~those in eqs.(\ref{eq:vthree1}) to (\ref{eq:vthree2}))
into the gluon-loop self-energy diagram. The reader may however
argue that we are ignoring the considerable number of $v_{4}$-vertex
insertions (i.e.~those in eqs.(\ref{eq:vfour1}) to
(\ref{eq:vfour2})) which give rise to corrections
to the heavy-quark self-energy of the same order in the velocity as those
considered above (see Fig.4). This is of course true, but it is very easy to
show that the full effect of all possible $v_{4}$-vertex insertions
into the diagrams of Fig.3 is just to multiply them by a factor
$(1/v_{4})^{m-1}$.

To demonstrate this result, consider the diagrams
of Fig.3. We can insert $n$
$v_{4}$-vertices only in two ways, namely, A: one at any of the
quark-gluon vertices and the remaining $n-1$ on the heavy-quark line and B:
all $n$ insertions into the heavy quark propagator inside the gluon loop.
The resulting diagrams are shown in Fig.4. Now, the
effect of a $v_{4}$-vertex insertion into a heavy quark propagator is simply to
multiply the same propagator by $(-)\, (v_{4} - 1)$, as it can be seen
from the Feynman rule in (\ref{eq:vfour1}). If the insertion is at a
quark-gluon vertex, the effect is to multiply the same vertex by $(v_{4}-1)$
(see eq.(\ref{eq:act2})).
For example,
two $v_{4}$-vertex insertions, one at the quark-gluon vertex and the other into
the heavy-quark propagator inside the gluon loop, give $(-)\, (v_{4}-1)^{2}$
times the old diagram without any $v_{4}$-vertex insertions.  Therefore,
we only have to count the number of topologically different diagrams with
$n$ $v_{4}$-vertex insertions in each class of Fig.4, for all diagrams in a
class give the same contribution to the heavy-quark self-energy. To this end,
we observe
that the number of different ways we can insert $n$ $v_{4}$-vertices on a
heavy-quark line where there are $m$ $v_{3}$-vertices
is $(n+m)!/[n!\, m!]$ (the old combinatorial problem of distributing
$n$ balls in $m+1$ boxes). Therefore, the sum of the graphs in Fig.4 gives
\begin{eqnarray}
{\rm Fig.4} &=& (-)^{n}\, \mfrac{(n+m-2)!}{[n!~(m-2)!]}\,
(v_{4}-1)^{n}\, {\rm C.1}\, +\,
(-)^{n}\, \mfrac{(n+m)!}{[n!~m!]}\, (v_{4}-1)^{n}\, {\rm C.2}\nonumber\\
&+& 2\, (-)^{n-1}\, \mfrac{(n-1+m)!}{[(n-1)!~m!]}\, (v_{4}-1)^{n}\, {\rm C.2}\,
+\,
(-)^{n-2}\, \mfrac{(n-2+m)!}{[(n-2)!~m!]}\, (v_{4}-1)^{n}\, {\rm
C.2}\nonumber\\
&=& (-)^{n}\, \mfrac{(n+m-2)!}{[n!~(m-2)!]}\, (v_{4}-1)^{n}\, ({\rm C.1}\, +\,
{\rm C.2} )
\end{eqnarray}
Summing from $n=0$ to $\infty$, we get
\be
\sum_{n=0}^{\infty}\, {\rm Fig.4} = \mfrac{1}{v_{4}^{m-1}}\, ({\rm C.1}\, +\,
{\rm C.2})
\ee
as anticipated.

We turn now to the computation of diagrams of Fig.3. Their amplitudes
are
\begin{eqnarray}
C_1(m,p) &=& \mfrac{-1}{a}\, \left(\mfrac{\alpha_{s}}{\pi}\right)\, C_{F}\,
(-v_{3})^{m}\, e^{-2 i p_{4}}\, \mfrac{1}{2 \pi}\, \int^{\pi}_{-\pi}\,
\mfrac{d^{3}k}{(2 \pi)^{3}}\, \sin^{m}(p_{3}-k_{3})\nonumber \\
&\times& \oint\, \mfrac{dz}{2 \pi i\, z}\, \mfrac{z}{[1\, +\, \epsilon\, -\,
e^{-i p_{4}}\, z]^{m+1}}\, \mfrac{-z}{(z-z_{-})(z-z_{+})}
\end{eqnarray}
and
\begin{eqnarray}
C_2(m,p) &=& \mfrac{1}{a}\, \left(\mfrac{\alpha_{s}}{\pi}\right)\, C_{F}\,
(-v_{3})^{m}\, \mfrac{1}{2 \pi}\, \int^{\pi}_{-\pi}\,
\mfrac{d^{3}k}{(2 \pi)^{3}}\, \sin^{m-2}(p_{3}-k_{3})\, \cos^{2}(p_{3}-k_{3}/2)
\nonumber\\
&\times& \oint\, \mfrac{dz}{2 \pi i\, z}\, \mfrac{z}{[1\, +\, \epsilon\, -\,
e^{-i p_{4}}\, z]^{m-1}}\, \mfrac{-z}{(z-z_{-})(z-z_{+})}
\end{eqnarray}
where $z=e^{i k_{4}}$, the contour integral is along the unit circle and we
have
used the fact that the gluon propagator can be written as
\be
\mfrac{1}{\sum_{\mu=1}^{4}\, (1-\cos(k_{\mu}))\, +\, (a \lambda)^{2}/2}\, =\,
\mfrac{-z}{(z-z_{-})(z-z_{+})}
\ee
with $z_{\pm}$ being the solutions of
$z_{\pm}^{2}\,-\,2\, (1+B)\, z_{\pm}\, +\, 1\,=\, 0$ and
\be
B=\sum_{\mu=1}^{3}\, (1-\cos(k_{\mu}))\, +\, (a \lambda)^{2}/2
\ee

The non-vanishing terms as $a$ goes to zero are $C_{1,2}(p=0)$, which contain a
linear divergence, and the first derivatives of $C_{1,2}(p)$ with respect to
$p_{3}$ and $p_{4}$ at $p=0$, which are logarithmically divergent. In either
case, the calculation reduces to the
computation of the contour integral over $z$ which can easily be performed
taking into account that only the pole
$z=z_{-}$ lies in the unit circle. Furthermore, the $\epsilon$-prescription
tells
us that the pole of the quark propagator does not contribute to the contour
integral when the Cauchy's theorem is used. The final result is, for the
non-derivative contribution,
\begin{eqnarray} \label{p1}
C_{1}(2m+1,p=0) &=& 0\; \;  \;   {\rm by}\, {\rm parity}\nonumber\\
C_{1}(2m,p=0) &=& \mfrac{-1}{a}\, \left(\mfrac{\alpha_{s}}{\pi}\right)\,
C_{F}\, v_{3}^{2 m}\, \mfrac{2}{4^{m+1}}\, \left[\, {\rm Si}^{(20)}(m-1)\, +\,
{\rm Si}^{(11)}(m-1)\, \right]\nonumber\\
C_{2}(2m+1,p=0) &=& 0\;  \;  \;  {\rm by}\, {\rm parity}\nonumber\\
C_{2}(2m,p=0) &=& \mfrac{1}{a}\, \left(\mfrac{\alpha_{s}}{\pi}\right)\,
C_{F}\, v_{3}^{2 m}\, \mfrac{1}{4^{m}}\, \left[\, {\rm Cs}^{(10)}(m-1)\, +\,
{\rm Cs}^{(01)}(m-1)\, \right]
\end{eqnarray}
The derivative with respect to $p_{3}$ at $p=0$ gives
\begin{eqnarray} \label{p2}
\left(\mfrac{\partial C_{1}(2m,p)}{\partial p_{3}}\right)_{0} &=& 0 \; \;  \;
{\rm by}\, {\rm parity}\nonumber\\
\left(\mfrac{\partial C_{1}(2m-1,p)}{\partial p_{3}}\right)_{0} &=&
\left(\mfrac{\alpha_{s}}{\pi}\right)\,
C_{F}\, v_{3}^{2 m -1}\, \mfrac{2 m -1}{4^{m}}\, \left[\, 2\,
{\rm Cs}^{(11)}(m-1)\, -\,
{\rm Id}^{(11)}(m-1)\, \right]\nonumber\\
\left(\mfrac{\partial C_{2}(2m,p)}{\partial p_{3}}\right)_{0} &=& 0\; \;  \;
{\rm by}\, {\rm parity}\nonumber\\
\left(\mfrac{\partial C_{2}(2m+1,p)}{\partial p_{3}}\right)_{0} &=&
\left(\mfrac{\alpha_{s}}{\pi}\right)\,
C_{F}\, v_{3}^{2 m + 1}\, \mfrac{1}{4^{m}}\nonumber\\
&\times& \left[\, \mfrac{2 m +1}{2}\,
\left\{ {\rm Si}^{(10)}(m-1)\, +\, {\rm Si}^{(01)}(m-1)\, +\,
{\rm Si}^{(11)}(m-1)\right\} \right.\nonumber\\
&-&\, (2 m -1)\, \left. \left\{ {\rm Cs}^{(10)}(m-1)\, +\, {\rm Cs}^{(01)}(m-1)
\, +\, {\rm Cs}^{(11)}(m-1)\right\} \right]
\end{eqnarray}
Finally, the derivatives with respect to $p_{4}$ at $p=0$ yield
\begin{eqnarray}  \label{p3}
\left(\mfrac{\partial C_{1}(2m+1,p)}{\partial p_{4}}\right)_{0} &=& 0\; \;  \;
{\rm by}\, {\rm parity}\nonumber\\
\left(\mfrac{\partial C_{1}(2m,p)}{\partial p_{4}}\right)_{0} &=& i\,
\left(\mfrac{\alpha_{s}}{\pi}\right)\,
C_{F}\, v_{3}^{2 m }\, \mfrac{1}{4^{m}}\nonumber\\
&\times& \left[\, {\rm Si}^{(20)}(m-1)\, +\,
{\rm Si}^{(11)}(m-1)\, +\, \mfrac{2 m +1}{2}\, {\rm Si}^{(21)}(m-1)\,
\right]\nonumber\\
\left(\mfrac{\partial C_{2}(2m+1,p)}{\partial p_{4}}\right)_{0} &=& 0\; \;  \;
{\rm by}\, {\rm parity}\nonumber\\
\left(\mfrac{\partial C_{2}(2m,p)}{\partial p_{4}}\right)_{0} &=&  -i\,
\left(\mfrac{\alpha_{s}}{\pi}\right)\,
C_{F}\, v_{3}^{2 m}\, \mfrac{2 m -1}{4^{m}}\, {\rm Cs}^{(11)}(m-1)
\end{eqnarray}
where ${\rm Si}^{(\alpha \beta)}(m)$, ${\rm Cs}^{(\alpha \beta)}(m)$  and
${\rm Id}^{(\alpha \beta)}(m)$ are three-dimensional integrals which
analytical expressions and numerical values for several $m$ can be found in
appendix A and Table A.1 respectively.

We are now in a position to giving the expression of the heavy-quark
self-energy on the lattice at any order in the velocity $v_{3}$. In fact,
it can be written as
\begin{eqnarray} \label{self}
\Sigma(p,v)&=& \mfrac{1}{a}\, \alpicf\, \, v_{4}\, \sum_{i=0}^{\infty}\,
\left(\mfrac{v_{3}}{v_{4}}\right)^{2i}\, \Sigma^{(2i)}_{0}\nonumber\\
&+&\, \alpicf\,
\left[\, \sum_{i=0}^{\infty}\,
\left(\mfrac{v_{3}}{v_{4}}\right)^{2i}\, \left\{ \Sigma^{(
2i)}_{40}\, -\, \Sigma^{(2i)}_{41}\, \log(a \lambda)\, \right\}\, \right]\,
( i p_{4} v_{4})\nonumber\\
&+&\, \alpicf\,
\left[\, \sum_{i=0}^{\infty}\,
\left(\mfrac{v_{3}}{v_{4}}\right)^{2i}\, \left\{ \Sigma^{(2i+1)}_{30}\, -\,
\Sigma^{(2i+1)}_{31}\, \log(a \lambda)\, \right\}\, \right]\,
(p_{3} v_{3})\nonumber\\
&+& O(\alpha_{s}\, a)
\end{eqnarray}
where the constants $\Sigma^{(m)}_{0,30,40}$ can easily be obtained from the
results in eq.(\ref{p1}) to (\ref{p3}). Their analytical expressions and
numerical values are listed in appendix A and Table A.2.

To finish this section,  we wish to discuss in detail the values of
constants $\Sigma^{(m)}_{41}$ and $\Sigma^{(m)}_{31}$ which determine both
the one-loop
wave function renormalization of the heavy-quark with velocity $v_{3}$ and
the anomalous dimension of the operator $\opv$.

We begin with $\Sigma^{(2m)}_{41}$. For $m=0$, it is the static heavy-quark
wave-function renormalization, first computed in ref.\cite{eihil2}. We
reproduce their result $\Sigma^{(0)}_{41}=4$ (see eq.(\ref{anaex2}) and
Table A.1).
If $m>0$,
$\Sigma^{(2m)}_{41}$ is determined by the pole part of the
sum of the derivatives with respect to $p_{4}$ of diagrams C.1 and C.2.
The terms that contain a logarithmic divergence are ${\rm Si}^{(21)}(m)$ and
${\rm Cs}^{(11)}(m)$ which appear in $\Sigma^{(2m)}_{41}$ through the
combination
\be \label{aa1}
\Sigma^{(2m)}_{41}\, \propto\, \mfrac{2 m+1}{2}\,
{\rm Si}^{(21)}(m)\mid_{\rm pole}\, -\, \mfrac{2 m -1}{2}\,
{\rm Cs}^{(11)}(m)\mid_{\rm pole}
\ee
In appendix B, we calculate the logarithmically divergent part of
${\rm Si}^{(21)}(m)$ and ${\rm Cs}^{(11)}(m)$. Inserting eqs.(\ref{apex1}) and
(\ref{apex11}) into
(\ref{aa1}), we obtain that these integrals conspire order by order in the
velocity to yield a vanishing coefficient $\Sigma^{(2m)}_{41}$ for $m>0$.
In other
words, the coefficient of the logarithm of the wave-function renormalization
of a heavy-quark moving with velocity $v_{3}$ is independent of the velocity,
as it should be. Moreover, the anomalous dimension turns out to be equal to
that of the static theory, which in turn is the same as the one in the
continuum. Therefore, we can say that the lagrangian (\ref{eq:act}) preserves
both the infrared and ultraviolet behaviour of the non-expanded theory order
by order in the velocity.

Similarly, the pole part of the sum of the derivatives with respect
to $p_{3}$ of diagrams C.1 and C.2, determine $\Sigma^{(2m+1)}_{31}$.
In this case, there is only a term containing a logarithmic divergence,
${\rm Cs}^{(11)}(m)$, which appear in $\Sigma^{(2m+1)}_{31}$ through the
combination
\be  \label{aa2}
\Sigma^{(2m+1)}_{31}\, \propto\, \mfrac{2 m+1}{4^{m+1}}\,
\left\{\, 2\, {\rm Cs}^{(11)}(m)\, -\, {\rm Id}^{(11)}(m)\,
\right\}\mid_{\rm pole}\, -\, \mfrac{2 m -1}{4^{m}}\,
{\rm Cs}^{(11)}(m-1)\mid_{\rm pole}
\ee
Again, substituting the expressions for the pole parts given in
eqs.(\ref{apex11}) and (\ref{apex12}) into (\ref{aa2}),
we observe that the coefficients of the logarithms conspire
order by order in the velocity
to produce a vanishing $\Sigma^{(2m+1)}_{31}$ for $m>0$. The only
logarithmically divergent term left is that for $m=0$. Note also that
$\Sigma^{(1)}_{31}$ is equal to $\Sigma^{(0)}_{41}$. This fact it very
important because it implies that the renormalization constant of the operator
$\opv$ is finite (see next section), as it should be since this operator
is conserved in the static theory. Again,
consistency with the non-expanded theory is explicitly shown.

Putting all these things together, we have
\begin{eqnarray} \label{self2}
\Sigma(p,v)&=& \mfrac{1}{a}\, \alpicf\, v_{4}\,
\sum_{i=0}^{\infty}\,
\left(\mfrac{v_{3}}{v_{4}}\right)^{2i}\, \Sigma^{(2i)}_{0}\nonumber\\
&+&\, \alpicf\, \left[\,
\sum_{i=0}^{\infty}\,\left\{ \,
\left(\mfrac{v_{3}}{v_{4}}\right)^{2i}\, \Sigma^{(2i)}_{40} \,
\right\}\, -\, 4\, \log(a \lambda)\, \right]\, ( i p_{4} v_{4})\nonumber\\
&+&\, \alpicf\, \left[\, \sum_{i=0}^{\infty}\,
\left\{\, \left(\mfrac{v_{3}}{v_{4}}\right)^{2i}\, \Sigma^{(2i+1)}_{30}
\, \, \right\} -\, 4\, \log(a \lambda)\, \right]\,
(p_{3} v_{3})\nonumber\\
&+& O(\alpha_{s}\, a)
\end{eqnarray}
which is one of our most important results.
Notice that up to $O(v_{3}^{2})$,
the heavy-quark self-energy (\ref{self2}) coincides (within an error of
less than 1\%) with the one computed in the previous
section using a different integration method (see eq.(\ref{oldsig})). This
fact makes us think that our numerical calculation is correct.

\subsection{Wave function and mass renormalization}

Having obtained the heavy-quark self-energy, we want to define and compute
its wave function renormalization $Z_{Q}$ (defined by
$Q^{R} = Z_{Q}^{-1/2} Q^{B}$ ) and mass renormalization $\delta M$.

In order to get the renormalization constants, we study the heavy-quark
propagator near on-shell including order $\alpha_{s}$ corrections
\be  \label{popro}
i H(p,v_{3})\, =\, \mfrac{1}{(1-\Sigma_{4})\, (i p_{4} v_{4})\, +
(1-\Sigma_{3})\, (p_{3} v_{3})\, -\, \Sigma_{0}\, +\, O(p^{2})}
\ee
where
\be
\Sigma_{4}\, =\, -i\, \left(\mfrac{\partial \Sigma(p)}{\partial(v_{4}p_{4})}
\right)(0)\;\;\;\;\;
\Sigma_{3}\, =\, \left(\mfrac{\partial
\Sigma(p)}{\partial(v_{3}p_{3})}\right)(0)
\ee
The analytical expressions for $\Sigma_{4}$ and $\Sigma_{3}$ are readily
computable from eq.(\ref{self2}).

If we impose on-shell renormalization conditions along with the
normalization of the velocity $v^{2}=1$, it is easy to check that
up to order $\alpha_{s}$ \cite{real}
\begin{eqnarray}
\label{pepo1}\delta M &=& - \Sigma_{0}\\
Z_{Q} &=& 1\, +\, v_{4}^{2}\, \Sigma_{4} - v_{3}^{2}\, \Sigma_{3}\nonumber\\
  &=& 1\, +\, \alpicf\, \left[\,- 4\, \log(a \lambda)\, +\, v_{4}^{2}\,
\Sigma^{(0)}_{40}\, +\,  v_{4}^{2}\, \sum_{i=0}^{\infty}\,
\left(\mfrac{v_{3}}{v_{4}}\right)^{2i}\, \left\{\, \Sigma^{(2i+2)}_{40} \,
-\, \Sigma^{(2i+1)}_{30} \, \right\}\, \right] \label{pepo}\\
Z_{v}^{-1} &=& 1\,+\, v_{4}^{2}\, \Sigma_{4}\, -\, v_{4}^{2}
\, \Sigma_{3}\nonumber\\
  &=& 1\, +\, \alpicf\, v_{4}^{2}\, \sum_{i=0}^{\infty}\,
\left(\mfrac{v_{3}}{v_{4}}\right)^{2i}\, \left\{\, \Sigma^{(2i)}_{40} \,
-\, \Sigma^{(2i+1)}_{30} \, \right\} \label{poppa}
\end{eqnarray}
where $Z_{v}$ is the renormalization of the heavy quark velocity, first
introduced in ref.\cite{real}, defined by
\be
v^{R}_{3}\, =\, Z_{v}^{-1}\,  v^{B}_{3}
\ee
As we will see in the next section, $Z_{v}$ is a lattice effect that
originates from the fact that the wave function and mass renormalizations
are not sufficient to match the lattice and continuum amplitudes. In this
sense, this 'velocity' renormalization can be interpreted also as the
matching constant necessary to reproduce the physical amplitudes
in the continuum from the ones on the lattice.
We will return to this subject in the next section.

For future use, it is convenient to write $Z_{Q}$ and $Z_{v}$ as
\begin{eqnarray}
Z_{Q} &=& \, 1\, +\, \alpicf\, \left[\, -4\, \log(a \lambda)\, +\,
v_{4}^{2}\, \sum_{i=0}^{\infty}\,
\left(\mfrac{v_{3}}{v_{4}}\right)^{2i}\, Z^{(2i)}_{Q}\, \right] \label{pipi}\\
Z_{v} &=& \, 1\, +\, \alpicf\, v_{4}^{2}\, \sum_{i=0}^{\infty}\,
\left(\mfrac{v_{3}}{v_{4}}\right)^{2i}\, Z^{(2i)}_{v}
\end{eqnarray}
and the reader can find the numerical values of $Z^{(m)}_{Q,v}$ in Table A.2.

To finish this section we wish to briefly comment on some particularly
interesting
characteristics of $Z_{Q}$ and $Z_{v}$. We observe
that $Z_{Q}$ is logarithmically divergent so that
the corresponding anomalous dimension coincides with the one in the
continuum and is independent of the velocity.
On the other hand, $Z_{v}$ is finite because the logarithmically divergent
terms coming from the self-energy cancel out exactly in eq.(\ref{poppa}) order
by order in the velocity.
Finally, since
lattice regularization breaks the $O(4)$ symmetry,
the renormalization constants of the effective theory depend on the velocity
of the heavy quark. This does not happen in a Lorentz invariant theory
because there all quantities must depend on Lorentz invariants.

\subsection{Renormalization and matching of $\opv$}

The 'velocity' operator $\opv$ does not
renormalize multiplicatively and, in general, we need to perform subtractions
of terms that diverge as powers of the ultra-violet cut-off $a^{-1}$.
Specifically, we will demonstrate that $\opv$ mixes under
renormalization with the operator $\mkappa(x)=Q^{\dagger}(x)D_4(x)Q(x)$
through a coefficient
free of power divergences and with the operator $\uno(x)=Q^{\dagger}(x)Q(x)$
whose coefficient diverges as $1/a$.

It is convenient to proceed order by order in the velocity because in this way
the matching can be understood better.

Consider a single insertion of $\opv$. The vanishing of the corresponding
amplitude at zero external momentum
(eqs.(\ref{eq:a3zero}) and (\ref{eq:a4zero})) imply that
$\opv$ does not mix with the operator $\uno(x)$ with a
linearly divergent coefficient (i.e. proportional to $1/a$).
There is only a multiplicative renormalization of $\opv$.
In order to obtain it, we need the
wave-function renormalization constant $Z$ of the heavy quark in the static
theory. It has been computed by many authors \cite{eihil2}
\beq\label{eq:zeta}
Z~=~1-i\left(\frac{\partial\Sigma}{\partial p_4}\right)_0~
=~1+\alpicf\,[-2\log(a\lambda)^2+\Sigma^{(0)}_{40}]
\eeq
where $\Sigma^{(0)}_{40}=24.48$ (see eq.(\ref{self2})). Then,
the one-loop matrix element of the bare operator $\opv$ between heavy quark
states is given by
\beqn\nonumber
\langle \opv \rangle&=&
 - \Bigg(1
-\frac{1}{v_{3}}\left(\frac{\partial A_3}{\partial p_3}\right)_0
-\frac{1}{v_{3}}\left(\frac{ \partial A_4 }{\partial p_3}\right)_0
-i\left(\frac{\partial \Sigma}{\partial p_4}\right)_0\Bigg)~(v_{3} p_3 )\,
+\, \ldots
\\  \label{eq:rinsing}
&=&-\Bigg(~1+\alpicf\,
[\Sigma^{(0)}_{40}-\Sigma^{(1)}_{30}]~\Bigg)~(v_{3} p_3 )\,+\,\ldots
\eeqn
The dots indicate terms which vanish as $p^2$ for $p\rightarrow 0$,
and therefore do not contribute to the on-shell renormalization.
As anticipated in the previous section, the ultraviolet divergence of the
vertex correction cancels the one of the
field renormalization constant $Z$, leaving a finite term. This occurs
because the operator $\opv$ is conserved in the static theory.

This result can easily be generalized for an odd number $2m+1$ of insertions
of the operator $\opv$. In fact, from (\ref{p1})
we learn that it does not mix with
the operator $\uno$ because the corresponding amplitude vanishes at $p=0$ due
to the spatial parity invariance of the theory.
Therefore, two-quark matrix elements of $\opv$ do not contain linearly
divergent terms proportional to $\uno$.
The same reasoning applies to the mixing with the operator
$\mkappa$. Thus, $\opv$ renormalizes multiplicatively
with a finite renormalization constant at any odd order in the velocity.

Let us consider now the renormalization of the double insertion of $\opv$.
The amplitudes at zero external momentum are now non-vanishing,
\beq
B_3(0)~+~B_4(0)~\neq~0,
\eeq
implying that there is a mixing of the double insertion of $\opv$
with the operator $\uno(x)$ with a linearly divergent coefficient.
There is also a mixing of the double insertion of $\opv$
with the operator $\mkappa(x)$, because
\beq
\left( \frac{\partial B_3}{\partial p_4}\right)_0
+\left( \frac{\partial B_4}{\partial p_4}\right)_0~
\neq~0
\eeq
The mixing is finite because the
logarithmic divergences of $(\partial B_3/\partial p_4)_0$ and
$(\partial B_4/\partial p_4)_0$ cancel each other.
This is true at any order in the velocity, as we demonstrated before.

We have therefore the one-loop result
\be
\langle Q^{\dag}\opv Q\, Q^{\dag}\opv Q \rangle \,
=\, \frac{1}{a}\, \alpicf\,~v_{3}^2~\Sigma^{(2)}_{0}\,
\langle \uno \rangle ~
+~\alpicf~v_{3}^2~\Sigma^{(2)}_{40}~
\langle \mkappa \rangle+\cdots
\label{matchito}
\ee
where the dots indicate terms which do not contribute to the
on-shell renormalization.

As before, these results can be extended to any order in the velocity by means
of eqs.(\ref{p1}) and (\ref{p3}). We only give
the final result for the one-loop renormalized lattice operator $\opv$
\begin{eqnarray} \label{match1}
[\opv]^{(1)}_{Latt} &=&\Bigg(~1 + \alpicf\, v_{4}^{2}\, v_{3}\,
\left[\, \sum_{i=0}^{\infty}\,
\left(\mfrac{v_{3}}{v_{4}}\right)^{2i}\,
\left\{\, \Sigma^{(2i+1)}_{30}\, -\,  \Sigma^{(2i)}_{40}\,\right\}\,
\right]\, \Bigg)\,
[\opv]^{(0)}\nonumber\\
&+& \mfrac{1}{a}\,\alpicf\, v_{4}\,
\left[\, \sum_{i=0}^{\infty}\,
\left(\mfrac{v_{3}}{v_{4}}\right)^{2i}\, \Sigma^{(2i)}_{0}\, \right]\,
[\uno]^{(0)}\nonumber\\
&+& \alpicf\,
\left[\, \sum_{i=0}^{\infty}\,
\left(\mfrac{v_{3}}{v_{4}}\right)^{2i}\, \Sigma^{(2i)}_{40}\, \right]\,
[\mkappa]^{(0)}
\end{eqnarray}
where the superscripts $(0)$ denote bare operators.

To proceed further, we match the amplitudes of the bare lattice SHQET
onto the amplitudes of the
$\overline{MS}$-renormalized HQET (i.e.~non expanded in $v_{3}$). To do this we
need to know the two-quark amplitude of the operator $\opv$ for an external
momentum configuration on-shell. By direct computation, we observe that
order by order in the
velocity $v_{3}$ the sum of the relevant loop diagrams with
insertions of $v_{3}$-vertices vanish. Of course, there is a physical reason
for this to happen: the operator $\opv$ is conserved in the continuum static
theory, therefore it does not get renormalized by interactions with gluons.
Thus, we can write
\be \label{match2}
[\opv]^{(1)}_{\overline{MS}}\, =\, [\opv]^{(0)}\, +\, O(\alpha_{s}^{2})
\ee
In other words, the one-loop wave function renormalization of a heavy quark
is independent of its velocity due to the fact that
dimensional regularization is a covariant regularization.

We can now perform the continuum-lattice matching by computing the ratio of
the continuum amplitude to the lattice one. From eq.(\ref{match1}) and
(\ref{match2}) we have that the physical operator $\opv$ is related to the
lattice bare one by
\begin{eqnarray} \label{match}
[\opv]_{\overline{MS}} &=&\Bigg(~1+\alpicf\, v_{4}^{2}\,
\left[\, \sum_{i=0}^{\infty}\,
\left(\mfrac{v_{3}}{v_{4}}\right)^{2i}\, \left\{\,
 \Sigma^{(2i+1)}_{30}\, -\, \Sigma^{(2i)}_{40}\,\right\}\, \right]~\Bigg)\,
[\opv]_{Latt}\nonumber\\
&-& \mfrac{1}{a}\,\alpicf\,
v_{4}\, \left[\, \sum_{i=0}^{\infty}\,
\left(\mfrac{v_{3}}{v_{4}}\right)^{2i}\, \Sigma^{(2i)}_{0}\, \right]\,
[\uno]_{Latt}\nonumber\\
&-& \alpicf\,
\left[\, \sum_{i=0}^{\infty}\,
\left(\mfrac{v_{3}}{v_{4}}\right)^{2i}\, \Sigma^{(2i)}_{40}\, \right]\,
[\mkappa]_{Latt}\nonumber\\
&\equiv& Z_{v}\, \left[\, [\opv]_{Latt}\, -\, \mfrac{c_{1}}{a}\,  [\uno]_{Latt}
\, -\, c_{2}\,  [\mkappa]_{Latt}\, \right]
\end{eqnarray}
with obvious notation. As anticipated, we learn from the previous
equation that $Z_{v}$ can be interpreted as the lattice-continuum matching
constant of the operator $\opv$.

An equivalent way of performing the matching is to expand the
$\overline{MS}$-renormalized propagator in the HQET and compare it order by
order in the $v_{3}$ with the
propagator in the SHEQT on the lattice.
In the continuum we have
\beqn
iS(k)&=&\frac{ Z_{\overline{MS}} }{iv_4k_4+v_{3}k_3}~
=~\frac{ Z_{\overline{MS}} }{ik_4+\epsilon}
 +\frac{ \sqrt{Z_{\overline{MS}}} }{ik_4+\epsilon}(-v_{3} k_3)
  \frac{ \sqrt{Z_{\overline{MS}}} }{ik_4+\epsilon}
\nonumber\\
&+& \frac{ \sqrt{Z_{\overline{MS}}} }{ik_4+\epsilon}(-v_3 k_3)
\frac{ 1 }{ik_4+\epsilon}(-v_3 k_3)
\frac{ \sqrt{Z_{\overline{MS}}} }{ik_4+\epsilon}
+\frac{ \sqrt{Z_{\overline{MS}}} }{ik_4+\epsilon}\, \frac{-iv_{3}^2}{2}\, k_4
\, \frac{ \sqrt{Z_{\overline{MS}}} }{ik_4+\epsilon}+\cdots
\label{eq:prop1}\eeqn
where $Z_{ \overline{MS} }$ is the heavy quark field renormalization
constant
\beq
Z_{ \overline{MS} }=1+\alpicf~2\log(\mu/\lambda)^2
\eeq
The bare lattice propagator in the SHQET is instead given
(near the mass-shell) by:
\beqn\nonumber
i\tilde{S}(k)&=&\frac{Z}{ik_4+\epsilon}
+\frac{ \sqrt{Z} }{ ik_4+\epsilon }
(-v_{3} k_3)~\Bigg(~1+\alpicf\,[\Sigma^{(0)}_{40}-\Sigma^{(1)}_{30}]~\Bigg)
\frac{ \sqrt{Z} }{ ik_4+\epsilon }
\\ \nonumber
&+&\frac{ \sqrt{Z} }{ ik_4+\epsilon }
(-v_{3}k_3)~\Bigg(~1+\alpicf\,[\Sigma^{(0)}_{40}-\Sigma^{(1)}_{30}]~\Bigg)
\frac{ 1 }{ ik_4+\epsilon }
(-v_{3}k_3)\nonumber\\
&\times&\Bigg(~1+\alpicf\, \left\{\, [\Sigma^{(0)}_{40}-\Sigma^{(1)}_{30}]~
+~\frac{1}{a}~v_{3}^2~\Sigma^{(2)}_{0}~
+~v_{3}^2~\Sigma^{(2)}_{40}~ik_4\, \right\}  \Bigg)
\frac{ \sqrt{Z} }{ ik_4+\epsilon }\nonumber\\
&+&\frac{ \sqrt{Z_{\overline{MS}}} }{ik_4+\epsilon}\, \frac{-iv_{3}^2}{2}\, k_4
\, \frac{ \sqrt{Z_{\overline{MS}}} }{ik_4+\epsilon}+\cdots
\label{eq:prop2}\eeqn
where $Z$ is the field renormalization constant of the static
lattice theory given in eq.(\ref{eq:zeta})
(we omit for simplicity the mass renormalization).

\noindent
Matching at lowest order in $v_3$ (static approximation) is realized
by introducing a matching constant $\zeta$
of the bare lattice regulated field
onto the $\overline{MS}$ renormalized field
\beq
Q_{ \overline{MS} }~=~\zeta~Q_{ L }
\eeq
where
\beq
\zeta~=~\frac{ Z_{\overline{MS}} }{ Z }~=~1+\alpicf\, [2\log(a\mu)^2-
\Sigma^{(0)}_{40}]
\eeq

\noindent
At order $v_{3}$, we must introduce a matching constant $Z_{v}$ defined by
\beq
\opv_{\overline{MS}}~=~Z_{v}~\opv_{BL}
\eeq
The comparison of eqs.(\ref{eq:prop1}) and (\ref{eq:prop2}) gives
\beq
Z_{v}~=~1+\alpicf\,[\Sigma^{(1)}_{30}-\Sigma^{(0)}_{40}]
\eeq

\noindent
Matching at order $v_{3}^2$ requires to subtract from the double insertion
of $\opv$, the contribution proportional to $\mkappa$
and the one proportional to $\uno$,
since they are absent in the HQET propagator. This means that there is a mixing
of these operators in the lattice-continuum matching.
Performing the subtraction above, we reproduce (\ref{matchito}).

\noindent
This procedure can be iterated to higher orders in the
velocity $v_{3}$ leading to eq.(\ref{match2}).

\section{Renormalization of the heavy quark current}

In this section, we deal with the renormalization of the heavy-quark current
\be
J(x)\, =\, Q^{\dag}(x,v)\, \Gamma\, Q(x,v')
\ee
describing the transition of a heavy quark with velocity $v$ into
a heavy quark of velocity $v'$. $\Gamma$ stands for any of the 16 Dirac
matrices.
We specialize our computation to the most interesting case $v'=(1,\vec{0})$
and $v=(0,0,v_{3},\sqrt{1+v_{3}^{2}})$.

We will demonstrate that the weak current $J(x)$ renormalizes multiplicatively
with a coefficient that is only logarithmically divergent. We will show
explicitly
that the one-loop anomalous dimension of $J(x)$ depends on the velocity and
coincides order by order in the velocity with the one computed within HQET
in the continuum.

For the sake of clarity, we divide this section in two parts.
In the first one, we compute the on-shell lattice matrix element of $J(x)$
between heavy quark states up to order $O(v_{3}^{2})$. The second
subsection is devoted to extend the previous result to all orders in the
velocity.
It is there where we will re-obtain the velocity-dependent
one-loop anomalous dimension by summing all the diagrams with insertions
of the operator $\opv$.

\subsection{Matrix element of the current up to $O(v_{3}^{2})$}

We start by considering the renormalization of the weak
current $J(x)$ with one insertion of the operator $\opv$. We study the
Green's function
\beq\label{eq:oj}
G(x,z)~=~\int \diff^4 y \diff^4 w\,
\langle 0\mid T~[ Q(x)\, Q(y)\opv Q^{\dagger}(y)\, J(w)\, Q^{\dagger}(z)]
 \mid 0\rangle
\eeq
The only non-vanishing diagram involved is drawn in Fig.5. The amplitude
of diagram D.1 is given by
\beq
D_1(p,q)~=~\frac{g^2C_Fv_{3}}{2}~e^{iq_4-2ip_4}
\int\frac{\diff ^4k}{(2\pi)^4}
\frac{ e^{-ik_4} \sin(k_3+p_3) }{ 1- e^{-i(k_4+p_4)} +i \epsilon }
\frac{1}{ \Delta_1(k) }
\eeq
where $p$ is the final momentum of $Q$ and $q$ is the
(incoming) momentum of $J$.
There is a potential logarithmic divergence, which is absent because the
integral vanishes by parity at zero external momenta
\beq
D_1(p=0,q=0)~=~0
\eeq
Therefore, there is not any additional renormalization of the Green's
functions of the form (\ref{eq:oj}).

Consider now the renormalization of the Green functions of $J(x)$ containing
a double insertion of $\opv$.
The relevant diagram is given in Fig.5.
The amplitude for D.2 is given, at zero external momenta, by
\beq
D_{2}~=~-\frac{g^2C_Fv^{2}_{3}}{6}~\int\frac{\diff^4k}{(2\pi)^4}
\frac{ \eta(\vec{k})~e^{-ik_4} }{ (1-e^{-ik_4}+\epsilon)^4 }
\frac{1}{\Delta_1(k)}
\eeq
With the technique described in sec.~5.1, the previous integral is
transformed into
\beqn\label{eq:a12}
D_{2}&=&\frac{g^2C_F}{16\pi^2}v^{2}_{3} \Bigg( -\frac{1}{36\pi^2}
\int d^4k \frac{ \eta(\vec{k}) }{ \Delta_1(k)^3 }
[~\cos k_4+2\cos 2k_4 +\cos 3k_4~]
\\ \nonumber
&-&\frac{1}{36\pi^2}
\int d^4k \frac{ \eta(\vec{k}) }{ \Delta_1(k)^2 }
[~3/2 + 3\cos k_4 + 2\cos 2k_4~]
-\frac{1}{12\pi}\int\diff^3k\frac{ \eta(\vec{k}) }{ \Delta_1(0,\vec{k})^2 }~
\Bigg)
\eeqn
The logarithmic singularity of the amplitude is entirely contained in
the first integral.
The numerical computation gives
\beq \label{ee2}
D_{2}~=~\alpicf\,~v_{3}^2\,
[~\frac{2}{3}\log(a\lambda)^2+ d_{2}~]
\eeq
where $d_{2}=-5.022$.

Adding to (\ref{ee2}) the contribution from the external wave-function
renormalization $Z_{Q}$, we obtain that the matrix element of the current
$J(x)$ between on-shell heavy-quark states is
\begin{eqnarray}
\langle c,v_{3}\mid J \mid b,\vec{0}\rangle &=& 1\, -\,
\left(\mfrac{\alpha_{s}}{\pi}\right)\,
\mfrac{C_{F}}{4}\, [\, Z_{\xi}^{(0)}\, +\, v_{3}^{2}\,
Z_{\xi}^{(2)}\, -\, \mfrac{4}{3}\, v_{3}^{2}\, \log(a \lambda)\, ]
\nonumber\\
&+& O(\alpha_{s}\, a\, , v_{3}^{3})
\label{yoyo}\end{eqnarray}
where $Z_{\xi}^{(0)}\, =\, -19.95$ is the old result for a static heavy quark
and $Z_{\xi}^{(2)} = -4.653$. The reason for the
introduction of the constants $Z^{(n)}_{\xi}$ will be apparent in section 6.3.

Note that the logarithmic divergence from the vertex diagram for two static
heavy quarks (lower order in $v_{3}$) exactly cancels
the one from the external wave-function renormalization resulting in a finite
lowest order correction to the matrix element of the current $Z^{(0)}_{\xi}$.
The physical
reason for this to happen is
that the flavour conserving current, i.e.~the current $J(x)$ for equal
velocities $v=v'$ or equivalently $v_{3}=0$, is conserved in the HQET and so
its anomalous dimension must be zero. Therefore, the anomalous dimension of
the current $J(x)$ starts from $v_{3}^{2}$ in an expansion in the velocity.

\subsection{Matrix element of the current beyond $O(v_{3}^{2})$}

The only non-vanishing diagram we need to calculate now has the same structure
as those in Fig.5 but with $m$ insertions of the operator $\opv$.
We will denote it by E.
All other possible one-particle irreducible diagrams vanish due to parity.
Again, we do not
consider the insertion of $v_{4}$-vertices because the net effect of all such
vertices is to multiply the original diagram by $1/v_{4}^{m}$, with $m$ the
number of $v_{3}$-vertices. The demonstration of this assertion is similar to
the case of the self-energy and so we do not repeat it here.

The amplitude corresponding to the diagram E is
\begin{eqnarray}
E(m,p) &=& - \left(\mfrac{\alpha_{s}}{\pi}\right)\, C_{F}\,
(-v_{3})^{m}\, e^{- 2 i p_{4}}\, \mfrac{1}{2 \pi}\, \int^{\pi}_{-\pi}\,
\mfrac{d^{3}k}{(2 \pi)^{3}}\, \sin^{m}(p_{3}-k_{3})\nonumber \\
&\times& \oint\, \mfrac{dz}{2 \pi i\, z}\, \mfrac{z}{[1\, +\, \epsilon\, -\,
e^{-i p_{4}}\, z]^{m+2}}\, \mfrac{-z}{(z-z_{-})(z-z_{+})}
\end{eqnarray}
where $z=e^{i k_{4}}$ and the contour integral is along the unit circle.

The non-vanishing term as $a$ goes to zero is $E(p=0)$, which contains a
linear divergence. First derivatives with respect to the external momentum give
rise to terms of order $O(a)$ that do not contribute to the on-shell
renormalization. The computation reduces to the
evaluation of the contour integral over $z$ which can be easily performed
taking into account that only the pole
$z=z_{-}$ lies in the unit circle. Furthermore, the $\epsilon$-prescription
tells
us that the pole of the quark propagator does not contribute to the contour
integral when the Cauchy's theorem is used. The result is,
\begin{eqnarray} \label{vp1}
E(2m+1,p=0) &=& 0\;\;\;  {\rm by}\, {\rm parity}\nonumber\\
E(2m,p=0) &=& - \left(\mfrac{\alpha_{s}}{\pi}\right)\,
C_{F}\, v_{3}^{2 m}\, \mfrac{2}{4^{m+1}}\nonumber\\
&\times& \left[\, {\rm Si}^{(20)}(m-1)\, +\,
{\rm Si}^{(11)}(m-1)\, +\, {\rm Si}^{(21)}(m-1)\right]
\end{eqnarray}
For $m=0$, the static case, the amplitude greatly simplifies
\be \label{puppa}
E(0,p=0)\, =\, - \left(\mfrac{\alpha_{s}}{\pi}\right)\,
\mfrac{C_{F}}{4}\, {\rm Id}^{(11)}(0)
\ee
The integrals ${\rm Si}^{(\alpha \beta)}(m)$ and ${\rm Id}^{(11)}(0)$
are defined and their numerical values tabulated in appendix A.

The matrix element of the current $J(x)$ between heavy-quark states is
logarithmically divergent for $m=0$  and also for $m>0$
because so is ${\rm Si}^{(20)}(m-1)$.  The coefficients of the logarithms
can easily be extracted order by order in the velocity from eqs.(\ref{apex1})
and (\ref{apex12}).
Adding the contribution from the wave-function renormalization of
the external states (see eq.(\ref{pepo}) and (\ref{pipi})), we have that
the matrix element of the current can be written as
\begin{eqnarray} \label{ass2}
\langle c,v_{3}\mid J \mid b,\vec{0}~\rangle \,=\,
1 &+& \left(\mfrac{\alpha_{s}}{\pi}\right)\,
\mfrac{C_{F}}{4}\, \sum_{i=0}^{\infty}\,
\, \left\{\, - v_{3}^{2i}\, Z^{(2i)}_{\xi}\right.\nonumber\\
&+& \left.\left(\mfrac{v_{3}}{v_{4}}\right)^{2i}\, 2\, \left[\,
\mfrac{1}{(2 i+1)}\, -\, \delta_{i,0}\, \right]\,
\log(a \lambda)^{2}\, \right\}
\,+\,O(\alpha_{s}\, a)
\end{eqnarray}
where $Z_{\xi}^{(m)}$ can easily be evaluated from eqs.(\ref{vp1}),
(\ref{puppa}) and (\ref{pipi}).
In order to simplify the computation in section 6.4 of
the relation between the Isgur-Wise function on the lattice
and in the continuum $\overline{MS}$, we have
expanded the finite contributions in powers of $v_{3}$ instead of
$v_{3}/v_{4}$. This is achieved by noting that
\be \label{zz}
\left(\mfrac{v_{3}}{v_{4}}\right)^{2m}\, =\, \sum_{j=0}^{\infty}\,
(-)^{j}\, \mfrac{(i+j-1)!}{j!\, (i-1)!}\, v_{3}^{2(i+j)}
\ee
For the numerical values of constants $Z^{(n)}_{\xi}$ we refer the reader to
Table A.2.

The only subtlety in this calculation is the fact that the wave function
renormalization at lowest order in the velocity (i.e.~the static case)
contributes to the matrix element with a coefficient twice the one
of higher velocity orders, which is $1/2$. The reason is that we consider
the $b$ quark static and the $c$ quark moving with a small velocity $v_{3}$.
Therefore, only the latter, as a
consequence of its interaction with the gluon field, gets
a velocity dependent wave function renormalization which lowest order is the
corresponding to a static heavy quark.

The interesting thing is that the infinite sum in front of $\log(a \lambda)$
can be evaluated simply recalling that
\be \label{ass}
\mfrac{1}{2 u}\, \log\left(\mfrac{1+u}{1-u}\right)\, =\,
\sum_{i=0}^{\infty}\, \mfrac{u^{2i}}{(2 i +1)}
\ee
Substituting (\ref{ass}) into (\ref{ass2}), we get
\begin{eqnarray} \label{current}
\langle c,v_{3}\mid J \mid b,\vec{0}~\rangle \,=\, 1 &+&
 \left(\mfrac{\alpha_{s}}{\pi}\right)\,
\mfrac{C_{F}}{4}\, \left[
\left(\, -2\, +\, \mfrac{v_4}{v_3}\,
\log\left(\mfrac{v_4+v_3}{v_4-v_3}\right)\, \right)\,
\log(a \lambda)^{2}\, -\,
\sum_{i=0}^{\infty}\, \left\{\, v_{3}^{2i}\,
Z_{\xi}^{(2i)}\, \right\}\, \right]
\nonumber\\
&+& O(\alpha_{s}\, a)
\end{eqnarray}
It should be stressed that the infrared structure of the matrix element
of the heavy-heavy current
is the same as the one evaluated within the continuum Georgi's
theory in ref.\cite{log}, as it should be. In other words, the expansion in the
velocity reproduces the correct infrared behaviour of the theory once we sum
all orders in the velocity. This is a check of the consistency of our approach.

Another check is provided by the comparison of the numerical value of the
constant $Z^{(2)}_{\xi}$ listed in Table A.2 (computed by direct contour
integration) and the one given below
eq.(\ref{yoyo}) (computed by integration by parts).
They coincide within error bars.

\subsection{Lattice-continuum matching}

In order to match the two-quark amplitude of the current to its counterpart
in the continuum, we need to compute the matrix element of the current in the
HQET renormalized in the $\overline{MS}$ scheme. This has already been done
by the authors of ref.\cite{log} and here we only quote their final result
\begin{eqnarray} \label{current2}
\langle c,v_{3}\mid J \mid b,\vec{0}~\rangle_{\overline{MS}}\,=\, 1 &+&
 \left(\mfrac{\alpha_{s}}{\pi}\right)\,
\mfrac{C_{F}}{4}\,
\left(\, 2\, -\, \mfrac{v_4}{v_3}\,
\log\left(\mfrac{v_4+v_3}{v_4-v_3}\right)\, \right)\,
\log(\mu/\lambda)^{2}\nonumber\\
&+& O(\alpha_{s}^{2})
\end{eqnarray}
where $\mu$ is the renormalization point.

By forming the ratio of (\ref{current2}) to (\ref{current}) we get the factor
that relates the matrix elements of the heavy-quark current $J(x)$ in the
lattice and in the continuum $\overline{MS}$ renormalization schemes
\begin{eqnarray} \label{matching}
\mfrac{\langle c,v_{3}\mid J \mid b,\vec{0}~\rangle_{\overline{MS}}}{
\langle c,v_{3}\mid J \mid b,\vec{0}~\rangle_{Latt}}\, =\,
1 &+& \left(\mfrac{\alpha_{s}}{\pi}\right)\,
\mfrac{C_{F}}{4}\, \left[\,
\left(\, -2\, +\, \mfrac{v_4}{v_3}\,
\log\left(\mfrac{v_4+v_3}{v_4-v_3}\right)\, \right)\,
\log(a \mu)^{2}\right.\nonumber\\
&+& \left.
\sum_{i=0}^{\infty}\, \left\{\, v_{3}^{2i}\,
Z_{\xi}^{(2i)}\, \right\}\, \right]\, +\, O(\alpha_{s}^{2})
\end{eqnarray}
As expected, the infrared regulator $\lambda$ disappears in the matching
because the lattice theory and its counterpart in the continuum have the same
infrared behaviour. It is the ultraviolet one that is different and
(\ref{matching}) takes this discrepancy into account.

Let us discuss now the on-shell renormalization of the lattice SHQET in the
real
space \cite{boucaud} instead of momentum space \cite{eihil2} as we have done
up to now. These renormalization schemes differ on the lattice and the
relation between them has been clarified in \cite{nonper}. A clear exposition
can be found in \cite{real} which we will follow almost verbatim here.

Consider the bare propagator of the heavy quark moving on the lattice with a
velocity $v_{3}$ along the z-axis as a function of time and momentum
$\vec{p}$. We will call it
$i H(t,\vec{p})$. This propagator can be obtained by
performing the Fourier transform with
respect to $p_{4}$, the fourth component of the external momentum, of the
propagator in the momentum space (\ref{popro}).
For large euclidean time and in the continuum limit,
$i H(t,\vec{p})$ reduces to
\be
i H(t,\vec{p})\, =\, Z_{Q}\, \mfrac{\theta(t)}{v^{R}_{4}}\,
\exp[-(t+1)/v^{R}_{4}\,\, (\delta M\, +\, v^{R}_{3}\cdot p_{3})\, ]
\ee
Note that for the momentum space renormalization conditions (\ref{pepo1}) to
(\ref{poppa}), $Z_{Q}$ is multiplied by an exponential with $(t+1)$
instead of $t$ in the heavy quark propagator.

On the other hand, in lattice simulations one fits the correlation
functions to an exponential evolution in euclidean time with $t$ instead
of $(t+1)$ (real space renormalization scheme). Therefore, if we do not
modify the momentum space renormalization
conditions appropriately, we will not take into account the correct wave
function
renormalization giving rise to a wrong lattice-continuum matching.
The solution is to take a shifted wave function
renormalization $\overline{Z}_{Q}$ related to the old one by \cite{eihil2}
\be
\overline{Z}_{Q}\, =\, Z_{Q}\, -\, \mfrac{\delta M}{v_{4}}
\ee
which tells us that he discrepancy between the momentum and the real space
renormalization schemes is finite and is given by the mass renormalization.
In Table A.2, we have tabulated the values of the renormalization
constants $\overline{Z}_{Q}$ and $\overline{Z}_{\xi}$ in the real space
lattice scheme.

\section{The Isgur-Wise function}

In this section we determine the relation between the value of the
derivatives of the Isgur-Wise
function at the zero recoil point, $\xi^{(n)}(1)$, measured on the lattice
and its physical counterparts in the continuum $\overline{MS}$.

In fact, taking $\mu=a^{-1}$ in (\ref{matching}), we find
\be \label{matcho}
\mfrac{\xi_{\overline{MS}}(v_{4})}{
\xi_{Lat}(v_{4})}\, =\,
1\, +\, \aalpicf\, \left\{\, Z_{\xi}^{(0)}\, +\, Z_{\xi}^{(2)}\,v_{3}^{2}\,
+\, Z_{\xi}^{(4)}\,v_{3}^{4}\, +\, \cdots\, \right\}
\ee
where the matching constants $Z^{(n)}_{\xi}$ are defined in eq.(\ref{ass2}).
Note that by setting $\mu=a^{-1}$ we have got ride of the
logarithms that appear in the lattice-continuum matching.

Substituting the expansions of both the Isgur-Wise function in the continuum
and the one on the lattice in powers of $v_{3}$ into (\ref{matcho}) and
demanding consistency order by order in $v_{3}$, we get
\begin{eqnarray} \label{xi}
\triangle \xi^{(1)}(a^{-1}) &=& \aalpicf\, \left\{\, Z^{(0)}_{\xi}\,
\xi^{(1)}_{Lat}(1)\, +\, 2\, Z^{(2)}_{\xi}\, \right\}\\ \label{xi1}
\triangle \xi^{(2)}(a^{-1}) &=& \aalpicf\, \left\{\, Z^{(0)}_{\xi}\,
\xi^{(2)}_{Lat}(1)\, +\, 4\, Z^{(2)}_{\xi}\, \xi^{(1)}_{Lat}(1)\right.
\nonumber\\
&+& \left. 2\, Z^{(2)}_{\xi}\, +\, 8\, Z^{(4)}_{\xi}\, \right\}\\
\label{xi2}
\triangle \xi^{(3)}(a^{-1}) &=& \aalpicf\, \left\{\, Z^{(0)}_{\xi}\,
\xi^{(3)}_{Lat}(1)\, +\, ( 6\, Z^{(2)}_{\xi}\, +\,
24\, Z^{(4)}_{\xi})\,
\xi^{(1)}_{Lat}(1)\right.\nonumber\\
&+&\, \left. 6\, Z^{(2)}_{\xi}\,
\xi^{(2)}_{Lat}(1)\, +\, 24\, Z^{(4)}_{\xi}\,
+\, 48\, Z^{(6)}_{\xi}\, \right\}
\end{eqnarray}
with $\xi^{(n)}(1)$ being the nth derivative of the Isgur-Wise function with
respect to $v_{4}$ at the zero recoil point $v_{4}=1$ and $\triangle
\xi^{(n)}(\mu_{0}) = \xi^{(n)}_{\overline{MS}}(1)\mid_{\mu=\mu_{0}} -
\xi^{(n)}_{Lat}(1)$.

\begin{table}[t] \label{tab0}\centering
\begin{tabular}{||c|rrrr||}
\hline
\hline
Isgur-Wise&\multicolumn{4}{c||}{Numerical Coefficients}\\
\cline{2-5}
derivatives&\multicolumn{1}{c}{$1$}&\multicolumn{1}{c}{$\xi^{(1)}_{Lat}(1)$}&
\multicolumn{1}{c}{$\xi^{(2)}_{Lat}(1)$}&
\multicolumn{1}{c||}{$\xi^{(3)}_{Lat}(1)$}\\
\hline
\hline
$\triangle \xi^{(1)}(a^{-1})$&$-9.31$&$-19.95$&$0.0$&$0.0$\\
$\triangle \xi^{(2)}(a^{-1})$&$3.44$&$-18.62$&$-19.95$&$0.0$\\
$\triangle \xi^{(3)}(a^{-1})$&$88.20$&$10.31$&$-27.92$&$-19.95$\\
\hline
$\triangle \overline{\xi}^{(1)}(a^{-1})$&$-24.88$&$0.0$&$0.0$&$0.0$\\
$\triangle \overline{\xi}^{(2)}(a^{-1})$&$17.35$&$-49.76$&$0.0$&$0.0$\\
$\triangle \overline{\xi}^{(3)}(a^{-1})$&$10.56$&$52.04$&$-74.64$&$0.0$\\
\hline
\hline
\end{tabular}
\caption{Numerical values of the constants determining the continuum--lattice
matching of the first derivatives of the Isgur-Wise function. A factor
$\alpicf$ multiplying all entries is understood.}
\end{table}

Eqs.(\ref{xi}) to (\ref{xi2}) are our most important results. They give the
one-loop relation at the scale $\mu=a^{-1}$ between the lattice measures of
the derivatives of
the Isgur-Wise function and their physical values. It should be stressed
that equivalent expressions can be obtain for the real space renormalization
scheme by replacing $Z^{(n)}_{\xi}$ by $\overline{Z}^{(n)}_{\xi}$.
In Table 1, we give the numerical values of the coefficients of
$\xi^{(n)}_{Lat}(1)$ in $\triangle \xi^{(n)}(a^{-1})$ both for
the momentum and real space (denoted with a bar) renormalization schemes.

As we mention in Sec.4, the values of the matching coefficients in Table 1
depend on the continuum renormalization scheme. Although we expect their
numerical values not to change very much in a different renormalization scheme,
consistency requires to properly include the contribution of the two-loop
anomalous dimension. In addition, we give the renormalization group evolution
of the derivatives of the Isgur-Wise function from the scale $a^{-1}$ to
a generic renormalization point $\mu$.

\begin{table}[t] \label{tab00}\centering
\begin{tabular}{||c|rrr||}
\hline
\hline
Isgur-Wise&\multicolumn{3}{c||}{R.~G.~correction}\\
\cline{2-4}
derivatives&\multicolumn{1}{c}{$a^{-1}=2$ GeV}&\multicolumn{1}{c}{$a^{-1}=4$
GeV}&\multicolumn{1}{c||}{$a^{-1}=6$ GeV}\\
\hline
\hline
$\triangle Z^{(1)}_{\xi}(\overline{m})$&$0.668$&$2.111$&$3.028$\\
$\triangle Z^{(2)}_{\xi}(\overline{m})$&$-0.248$&$-0.789$&$-1.135$\\
$\triangle Z^{(3)}_{\xi}(\overline{m})$&$0.155$&$0.489$&$0.701$\\
\hline
\hline
\end{tabular}
\caption{Renormalization group (R.~G.~) corrections
to the constants determining the
continuum--lattice
matching of the first derivatives of the
Isgur-Wise function at the scale $\mu=\overline{m}$ for several lattice
spacings $a$. The two-loop anomalous
dimension of the heavy-heavy quark current has been properly included.
A factor $\alpicf$ multiplying all entries is understood.}
\end{table}

The one-loop anomalous dimension $\gamma_{1}$ of the heavy-heavy quark
current is velocity dependent and has been calculated in eq.(\ref{current}).
On the other hand, the two-loop anomalous dimension $\gamma_{2}$ has been
computed in ref.\cite{hehev2} in the $\overline{MS}$ scheme.
Expanding both $\gamma_{1}$ and $\gamma_{2}$ as a power series in $v_{3}^{2}$
and inserting the result in eq.(\ref{xx3}), we find that the renormalization
group corrections $\triangle Z^{(n)}_{\xi}(\mu)$ to the
matching constants $Z^{(n)}_{\xi}$ are
\begin{eqnarray} \label{poppi}
\aalpicf\, \triangle Z^{(2)}_{\xi}(\mu) &=& - \mfrac{C_{F}}{3}
\,\mfrac{1}{\beta_{1}}\,
\left\{\, \log(\alpha_{s}(\mu)/\alpha_{s}(a^{-1}))\right.\nonumber\\
&+&\left(\mfrac{\alpha_{s}(a^{-1})}{\pi}\, -\,
\mfrac{\alpha_{s}(\mu)}{\pi}\right)\,
\left. \left[\, 2 \pi^{2}\,
+\, \mfrac{5}{18}\, N_{F}\, -\, \mfrac{29}{6}\, +\,
\mfrac{\beta_{2}}{\beta_{1}}\, \right]\, \right\}\\
\aalpicf\,\triangle Z^{(4)}_{\xi}(\mu)\, &=& - \mfrac{C_{F}}{3}
\,\mfrac{1}{\beta_{1}}\,
\left\{\, -\mfrac{2}{5}\,
\log(\alpha_{s}(\mu)/\alpha_{s}(a^{-1}))\right.\nonumber\\
&+& \left(\mfrac{\alpha_{s}(a^{-1})}{\pi}\, -\,
\mfrac{\alpha_{s}(\mu)}{\pi}\right)\,
\left. \left[\, -\mfrac{4 \pi^{2}}{5}
\, -\, \mfrac{1}{9}\, N_{F}\, +\, \mfrac{61}{40}\, -\,
\mfrac{2}{5}\, \mfrac{\beta_{2}}{\beta_{1}}\, \right]\, \right\}\\
\aalpicf\,\triangle Z^{(6)}_{\xi}(\mu)\, &=& - \mfrac{C_{F}}{3}
\,\mfrac{1}{\beta_{1}}\,
\left\{\,\mfrac{8}{35}\, \log(\alpha_{s}(\mu)/\alpha_{s}(a^{-1}))\right.
\nonumber\\
&+& \left(\mfrac{\alpha_{s}(a^{-1})}{\pi}\,-\
\mfrac{\alpha_{s}(\mu)}{\pi}\right)\,
\left. \left[\, \mfrac{\pi^{2}}{35}
\, +\, \mfrac{4}{63}\, N_{F}\, -\, \mfrac{43501}{37800}\, +\,
\mfrac{8}{35}\, \mfrac{\beta_{2}}{\beta_{1}}\, \right]\, \right\}
\end{eqnarray}
In Table 2, we report the numerical renormalization group corrections
to the constants $Z^{(n)}_{\xi}$ for several values of $a^{-1}$ at
the physical meaningful scale
$\mu=\overline{m}$ where $\overline{m}=
m_{B} m_{D}/(m_{B}+m_{D})\approx 1.4$ GeV, i.e.~the reduced mass of the
B- and D-meson. The number of active quarks is 3 because both the $b$ and the
$c$ quarks are taken to be static sources of color.
The values in Table 2 must be added to those of Table A.1 to obtain the
matching
constants at the scale $\overline{m}$.

\section{Power divergences}

In this section we prove that the renormalization
of $\xi^{(1)}(1)$ is not affected by ultraviolet power divergences.
The argument presented does not rely on any perturbative expansion
and is based only on the symmetries of the lattice $SHQET$.

A given operator $O$ can mix with lower and equal dimensional
operators $O'$ allowed by the symmetries of the (regulated) theory.
The mixing coefficients are proportional to the appropriate power
of the ultraviolet cut-off $1/a$ to account for the dimension.
If the dimension of $O$ and of $O'$ are the same, the mixing coefficients
contain in general logarithmic divergences, of the form $\log a$.
The computation of $\xi^{(1)}(1)$ involves single and double insertions
of the velocity operator $\opv=Q^{\dagger}(\vec{v}\cdot \vec{D})Q$.
The only possible linear divergence  in $\opv$ is through the mixing
with the operator $\uno=Q^{\dagger}Q$
\beq
<\opv>~=~\frac{k}{a}<\uno> +~{\it at~most~logarithmic~terms}
\eeq
where $k$ is a coefficient which has a perturbative expansion in $\alpha_s$.
Such a mixing is however impossible because of the spatial parity invariance
of the theory. Then, we have $k=0$.

The double insertion of $\opv$ can also mix with $\uno$, and in this case
the mixing is not forbidden by any symmetry
\beq
<\opv~\opv>~=~\frac{c}{a}~<\uno>  +~{\it at~most~logarithmic~terms}
\eeq
where now $c\neq 0$ in general. In sec.5 we have checked this result
with an explicit one-loop computation.

Therefore, the two- and three-point correlators defined in eqs.(\ref{eq:3pt})
and (\ref{eq:2D}) respectively, can be written as
\beqn\label{eq:c3s}
C_3^{(2)}(t,t')&=&\frac{c}{a}~(t'-t)~C_3^{(0)}(t,t')~
+~{\it at~most~logarithmic~terms}
\nonumber \\
C_D^{(2)}(t'-t)&=&\frac{c}{a}~(t'-t)~C_D^{(0)}(t'-t)~
+~{\it at~most~logarithmic~terms}
\eeqn
where the superscript indicates the order in the velocity.
The derivative of the Isgur-Wise function is given by the following
combination of correlation functions (see eq.(\ref{ratios1}))
\beq
\frac{C_3^{(2)}(t,t')}{C_3^{(0)}(t,t')}-\frac{C_D^{(2)}(t'-t)}{C_D^{(0)}(t'-t)}
\eeq
Substituting eqs.(\ref{eq:c3s}),
we see that the linear divergence cancel in
the expression for the $\xi^{(1)}(1)$, as anticipated.
We notice that the argument given is non-perturbative, and it is confirmed
by the explicit one-loop computations of the previous sections.

This argument, however, does not hold for higher derivatives of the Isgur-Wise
function. In fact, consider for example, the second derivative of this
function with respect to $v_{4}$. In this case, we must deal with
four insertions of the operator $\opv$. Therefore, the correlation functions
$C_{2,3}^{(4)}$ will contain a linearly divergent contribution just as
$C_{2,3}^{(2)}$. From eq.(\ref{ratios2}) we see that the correlation functions
$C_{2,3}^{(4)}$ enter the expression for $\xi^{(2)}(1)$ through the combination
\be \label{pippo2}
C_{3}^{(4)}\, C_{2}^{(0)}\, -\,  C_{3}^{(0)}\, C_{2}^{(4)}
\ee
which again is at most logarithmically divergent because the poles $1/a$ cancel
out. There is, however, a second contribution to $\xi^{(2)}(1)$ that is not
free
from linear divergences, namely,
\be \label{pippo3}
(C_{2}^{(2)})^{2}\,  C_{3}^{(0)}\,-\,
C_{3}^{(2)}\, C_{2}^{(2)}\, C_{2}^{(0)}
\ee
Substituting (\ref{eq:c3s}) into (\ref{pippo3}) we obtain that the term
$1/a^{2}$ cancels out in this combination but that proportional to $1/a$
survives giving rise to a linear divergence as $a$ goes to zero. Therefore,
the computation of $\xi^{(2)}(1)$ requires the subtraction both from
$C_{2}^{(2)}$ and $C_{3}^{(2)}$ of a linear
divergence as in the case of the self-energy of a quark.

The same reasoning can be applied to higher derivatives of the Isgur-Wise
function. As we increase the order of the derivative, the power of the
divergence also increases and thus non-perturbative subtractions from the
correlation functions of lower velocity degree are
necessary to obtain reliable results from a numerical computation on the
lattice.

\bc
\section{Conclusions}
\ec

We have studied the lattice renormalization of the
effective theory for slow heavy quarks,
which allows to compute the slope of the
Isgur-Wise function at the normalization point, $\xi^{(1)}(1)$, with
Montecarlo simulations.
We showed that the lattice-continuum renormalization constant
of $\xi^{(1)}(1)$ does not contain any linear ultraviolet divergence, but only
a logarithmic one. This implies that the matching of $\xi^{(1)}(1)$ can be done
in perturbation theory and it is not necessary to perform any
non-perturbative subtraction.
The lattice computation of the slope of the Isgur-Wise function
using the effective theory for
slow heavy quarks is therefore feasible in principle.

The one-loop lattice renormalization
constants of the slow heavy quark effective theory
have been computed to order $v^2$ together
with the matching constant of $\xi^{(1)}(1)$, which relates
the value of this form factor measured on the lattice to its
physical counterpart in the continuum.

We have demonstrated that the effective theory for slow heavy quarks
reproduces the infrared behaviour of the original (non-expanded) theory
order by order in the  velocity.
This means that we are dealing with a consistent expansion
of the effective theory for heavy quarks.

We have analysed the lattice effective theory for slow heavy quarks also
to higher orders in the velocity.
Unfortunately, the renormalization of the
higher derivatives of the Isgur-Wise function,
$\xi^{(n)}(1)$ for $n>1$, is affected by ultraviolet power divergences.
The lattice-continuum matching of $\xi^{(n)}(1)$ is therefore much
more involved than in the case of $\xi^{(1)}(1)$.
We stress however that the higher derivatives of the Isgur-Wise
function are much less important than the first one.
The renormalization problems of the slow heavy quark effective theory
which arise in higher orders do not constitute therefore a serious
limitation for its phenomenological applications.

We hope that the results of our analysis may encourage
the scientific community to
carry out the numerical simulation of $\xi^{(1)}(1)$ using the
slow heavy quark effective theory.
We believe that this theory can be a
source of interesting physical results.

\bc
\section*{Acknowlegments}
\ec

We are specially grateful to Prof.~G.~Martinelli for suggesting this
problem to us and for helpful discussions.
V.G.~wishes to thank the Istituto di Fisica ''G. Marconi'' of the
Universit\`a di Roma ''La Sapienza'' for its hospitality.
U.A.~wishes to thank the  Department of Physics
of Caltech, where the work has been completed, for its hospitality.
The work of V.G.~was supported in part by funds provided by CICYT under Grant
AEN 93-0234.
V.G.~acknowledges also the European Union for their support through the award
of a Postdoctoral Fellowship.
\newpage

\noindent
\section*{Table captions}

\noindent
{\bf Table 1}: Numerical values of the constants determining the
continuum--lattice matching of the first derivatives of the Isgur-Wise
function.
\vskip 0.4cm
\noindent
{\bf Table 2}: Renormalization group (R.~G.~) corrections
to the constants determining the
continuum--lattice
matching of the first derivatives of the
Isgur-Wise function at the scale $\mu=\overline{m}$ for several lattice
spacings $a$. The two-loop anomalous
dimension of the heavy-heavy quark current has been properly included.
A factor $\alpicf$ multiplying all entries is understood.
\vskip 0.4cm
\noindent
{\bf Table A.1}: Numerical values of three-dimensional integrals
for several values of $m$, the order in the expansion in powers
of the velocity $v_{3}$.
\vskip 0.4cm
\noindent
{\bf Table A.2}: Numerical values of the constants entering the
continuum--lattice
matching of the heavy-quark current for several values of $m$,
the order in the velocity expansion.
\vskip 0.4cm
\noindent

\noindent
\section*{Figure captions}

\noindent
{\bf Figure 1}: The diagrams contributing to the one-loop heavy quark
self-energy with one insertion of $\opv$, denoted by a crossed circle.
The double line represents
the heavy quark with velocity $(0,0,v_{3})$.
\vskip 0.4cm
\noindent
{\bf Figure 2}: The diagrams contributing to the one-loop heavy quark
self-energy with two insertions of $\opv$, denoted by a crossed circle.
The double line represents
the heavy quark with velocity $(0,0,v_{3})$.
\vskip 0.4cm
{\bf Figure 3}: The non-vanishing diagrams contributing to the one-loop
heavy quark self-energy with $m$ insertions of $\opv$, denoted by a crossed
circle.
The double line represents
the heavy quark with velocity $(0,0,v_{3})$.
\vskip 0.4cm
{\bf Figure 4}: The non-vanishing diagrams contributing to the one-loop
heavy quark
self-energy with $m$ insertions of $\opv$, denoted by a crossed circle, and
$n$ insertions of the operator $D_{4}$, represented by a full circle.
The double line represents
the heavy quark with velocity $(0,0,v_{3})$.
\vskip 0.4cm
{\bf Figure 5}: The non-vanishing one-particle irreducible
diagrams contributing to the one-loop vertex
of the current $J(x)$ with one and two insertions of $\opv$,
denoted by a crossed circle. The double line represents
the heavy quark with velocity $(0,0,v_{3})$.
The full line stands for a static heavy quark.
\vskip 0.4cm
\newpage

\begin{table}[p] \label{tab1}\centering
\begin{tabular}{||c|c|rrr|c||}
\hline
\hline
\mbox{}&\mbox{}&\multicolumn{3}{c|}{Velocity Power}&\multicolumn{1}{c||}{Infr.
Diver.}\\
\cline{3-6}
Type&B-Factor&$m=0$&$m=1$&$m=2$&$\log(a \lambda)$\\
\hline
\hline
Si&$(10)$&$8.284$&$26.148$&$178.85$&$0$\\
Si&$(11)$&$3.367$&$15.20$&$78.8$&$0$\\
Si&$(01)$&$5.791$&$24.877$&$120.76$&$0$\\
Si&$(20)$&$6.771$&$29.2$&$148$&$0$\\
Si&$(21)$&$-0.036$&$9.83$&$78.9$&$\mfrac{2\, 4^{m+1}}{2m+3}$\\
\hline
Ci&$(10)$&$13.34$&$34.8$&$149$&$0$\\
Ci&$(11)$&$2.485$&$12.93$&$72.9$&$\mfrac{4^{m+1}}{2m+1}$\\
Ci&$(01)$&$7.298$&$21.21$&$88.0$&$0$\\
\hline
Id&$(10)$&$19.95$&$39.2$&$241$&$0$\\
Id&$(11)$&$4.526$&$20.21$&$108.5$&$\mfrac{4^{m+1}}{2m+1}$\\
Id&$(01)$&$12.23$&$34.9$&$152$&$0$\\
\hline
\hline
\end{tabular}
\setcounter{table}{0}
\def\thetable{A.1}
\caption{Numerical values of three-dimensional integrals for several
values of $m$, the order in the expansion in powers
of the velocity $v_{3}$.}
\end{table}

\begin{table}[p] \label{tab2}\centering
\begin{tabular}{||c|ccccccc||}
\hline
\hline
\mbox{}&\multicolumn{7}{c||}{Velocity Power}\\
\cline{2-8}
Constant&$m=0$&$m=1$&$m=2$&$m=3$&$m=4$&$m=5$&$m=6$\\
\hline
\hline
$\Sigma^{(m)}_{0}$&$-19.95$&$0.0$&$15.57$&$0.0$&$8.20$&$0.0$&$7.75$\\
$\Sigma^{(m)}_{40}$&$24.48$&$0.0$&$7.60$&$0.0$&$7.55$&$0.0$&$8.65$\\
$\Sigma^{(m)}_{30}$&$0.0$&$12.67$&$0.0$&$7.28$&$0.0$&$8.37$&$0.0$\\
\hline
$Z^{(m)}_{Q}$&$24.48$&$0.0$&$-5.07$&$0.0$&$0.27$&$0.0$&$0.28$\\
$\overline{Z}^{(m)}_{Q}$&$4.53$&$0.0$&$30.45$&$0.0$&$-7.10$&$0.0$&$-0.17$\\
$Z^{(m)}_{v}$&$11.81$&$0.0$&$0.32$&$0.0$&$-0.82$&$0.0$&$--$\\
$Z_{\xi}^{(m)}$&$19.95$&$0.0$&$4.65$&$0.0$&$-1.59$&$0.0$&$-1.04$\\
$\overline{Z}_{\xi}^{(m)}$&$0.0$&$0.0$&$12.44$&$0.0$&$-5.28$&$0.0$&$2.42$\\
\hline
\hline
\end{tabular}
\setcounter{table}{0}
\def\thetable{A.2}
\caption{Numerical values of the constants entering the continuum--lattice
matching of the heavy-quark current for several values of $m$,
the order in the velocity expansion.}
\end{table}

\vfill
\eject

\myappendix

\bc
\appesection{Analytical expressions and numerical values of loop integrals}
\ec

The renormalization of both the heavy-quark current and the SHQET lagrangian
can
be written in terms of a few three-dimensional one-loop integrals. In this
appendix, we give their analytical expressions and numerical values up
to $O(v_{3}^{4})$.

We define
\begin{eqnarray}  \label{defsin}
{\rm Si}^{(\alpha \beta)}(m) &=& \mfrac{1}{2 \pi}\, \int^{\pi}_{-\pi}\,
d^{3}k\,\mfrac{\sin^{2}(k_{3})}{B^{\alpha}\,
\sqrt{[(2+B)B]^{\beta}}}\,\, \Xi(B)^{m}\nonumber\\
{\rm Cs}^{(\alpha \beta)}(m) &=& \mfrac{1}{2 \pi}\, \int^{\pi}_{-\pi}\,
d^{3}k\, \mfrac{\cos^{2}(k_{3}/2)}{B^{\alpha}\,
\sqrt{[(2+B)B]^{\beta}}}\,\, \Xi(B)^{m}\nonumber\\
{\rm Id}^{(\alpha \beta)}(m) &=& \mfrac{1}{2 \pi}\, \int^{\pi}_{-\pi}\,
d^{3}k\, \mfrac{1}{B^{\alpha}\,
\sqrt{[(2+B)B]^{\beta}}}\,\, \Xi(B)^{m}
\end{eqnarray}
where for the present calculation $\alpha=0,2$ and $\beta=0,1$, and
\be
\Xi(B)\,=\, \mfrac{\sin^{2}(k_{3})\, (2+B)}{B}\, \left[\, 1\, +\,
\mfrac{B}{\sqrt{(2+B)B}}\, \right]^{2}\,
=\, \mfrac{4\, \sin^{2}(k_{3})}{(1\,-\,z_{-})^{2}}
\ee
with $z_{-}$ the solution of
$z_{-}^{2}\,-\,2\, (1+B)\, z_{-}\, +\, 1\,=\, 0$ with $\mid z_{-}\mid < 1$.
Note that the function $\Xi(B)$ is infrared convergent. This fact will be used
in appendix B to subtract the infrared divergent behaviour from the integrals
in eq.(\ref{defsin}).

Other integrals as, for example, the one with a factor $\cos(k_{3})$ instead of
$\cos^{2}(k_{3}/2)$ can trivially be reduced to linear combinations of the
integrals defined in eq.(\ref{defsin}).

Obviously, these integrals must be evaluated numerically.
However, care should be taken when infrared divergences appear as
in ${\rm Si}^{(21)}(m)$,
${\rm Cs}^{(11)}(m)$ and ${\rm Id}^{(11)}(m)$.
In fact, in this case we cannot take $\lambda=0$.
The logarithmic infrared divergence must be
subtracted before computing them numerically. This has been done for arbitrary
$m$ in appendix B. We refer the reader to this appendix for details.
Other integrals are infrared finite for any value
of $m$ and thus can safely be computed by means of, for example, a Monte Carlo
routine.

Now, we give the analytical expressions of the heavy-quark self-energy
(see eq.(\ref{self2}))
\begin{eqnarray} \label{anaex}
\Sigma^{(2m)}_{0} &=& \mfrac{1}{4^{m-1}}\, \left[\,
{\rm Cs}^{(10)}(m-1)\, +\,
{\rm Cs}^{(01)}(m-1)\right.\nonumber\\
&-& \left. \mfrac{1}{2}\, {\rm Si}^{(20)}(m-1)\, -\, \mfrac{1}{2}\,
{\rm Si}^{(11)}(m-1)\, \right]\\
\Sigma^{(2m+1)}_{30} &=& \mfrac{1}{4^{m-1}}\,
\left[\, \mfrac{2 m +1}{2}\,
\left\{ {\rm Si}^{(10)}(m-1)\, +\, {\rm Si}^{(01)}(m-1)\, +\,
{\rm Si}^{(11)}(m-1)\right\}\, \theta(m) \right.\nonumber\\
&-&\, (2 m -1)\, \left\{ {\rm Cs}^{(10)}(m-1)\, +\, {\rm Cs}^{(01)}(m-1)
\, +\, {\rm Cs}^{(11)}(m-1)\right\}\, \theta(m)\nonumber\\
&+& \mfrac{2 m +1}{4}\, \left. \left\{ 2\, {\rm Cs}^{(11)}(m)\, -\,
{\rm Id}^{(11)}(m) \right\}\, +\, \mfrac{1}{4}\, {\rm Id}^{(01)}(m)\,
\delta_{m,0}\, \right]\\
\Sigma^{(2m)}_{40} &=&
\mfrac{1}{4^{m-1}}\, \left[\, {\rm Si}^{(20)}(m-1)\, +\,
{\rm Si}^{(11)}(m-1)\right.\nonumber\\
&+& \mfrac{2 m +1}{2}\, \left. {\rm Si}^{(21)}(m-1)\, -\, ( 2 m-1)\,
{\rm Cs}^{(11)}(m-1)\right]
\end{eqnarray}
which are supplemented with the old results for a static heavy quark
\be \label{anaex2}
\Sigma^{(0)}_{0}\,=\, -{\rm Id}^{(10)}(0)\;\;\;\;\;\;\;\;
\Sigma^{(0)}_{40} \,=\, {\rm Id}^{(10)}(0)\, +\,
{\rm Id}^{(11)}(0)
\ee

In table A.1, we list the numerical values of the three-dimensional
lattice regularized integrals (\ref{defsin}). These quantities has been
evaluated using both a
Monte Carlo and a lattice integration routine.
Errors are at most $O(1)$ in the last decimal place.

In table A.2, we present the numerical values for the heavy-quark
self-energy, the wave function renormalization, the velocity renormalization
and the lattice-continuum matching constants
for the Isgur-Wise function. As before, errors are at most $O(1)$ in the last
decimal place. The constants with a bar are in the real space lattice
renormalization scheme while the others has been calculated in the momentum
space scheme.

\bc
\appesection{Infrared subtraction of loop integrals}
\ec

Feynman integrals appearing in lattice perturbation theory must be
evaluated numerically because they are too much complicate to be integrated
analytically. The trouble arises when these integrals
are divergent as $\lambda$, the fictitious gluon mass, goes to zero. In order
to compute divergent lattice integrals, one has to subtract from the integrand
an expression which has its same infrared behavior \cite{twoloop2}.
Doing so, the integral to
be calculated can be expressed as a sum of an infrared finite one,
in which we can safely set $\lambda=0$, and a second integral which contains
the
divergences of the original one. The former can be evaluated numerically
while the later must be computed analytically to explicitly display the terms
which depend on the infrared and ultraviolet regulators.

In this calculation, all four-dimensional integrals can be reduced to
three-dimensional ones by integrating over the zeroth component of the loop
momentum or performing an integration by parts.
This simplification is possible because of the simple structure of
the heavy quark
propagator that only depends on the zeroth component of the momentum.
The integrands of the resulting three-dimensional integrals turn out to be
algebraic functions of B, defined by
\be
B\, =\, \sum_{\alpha=1}^{3}\, (1\,-\,\cos(k_{\alpha}))\, +\,
\mfrac{\lambda^{2} a^{2}}{2}
\ee
Therefore, it is convenient to know the infrared limit of B itself and some
other
functions of it. For $\mid k_{\alpha}\mid \ll 1$ we have
\begin{eqnarray} \label{expanb}
B  &\approx& \mfrac{1}{2}\, (\vec{k}^{2}\, +\, \lambda^{2})\, [\, 1\,-\,
\mfrac{1}{12}\, (\vec{k}^{2}\, +\, \lambda^{2})\,]\nonumber\\
\mfrac{1}{\sqrt{(1+B)^{2}\,-\,1}} &\approx& \mfrac{1}{\sqrt{\vec{k}^{2}\,
+\, \lambda^{2}}}\,[\, 1\,-\,
\mfrac{1}{12}\, (\vec{k}^{2}\, +\, \lambda^{2})\,]\nonumber\\
\mfrac{1}{B\,\sqrt{(1+B)^{2}\,-\,1}} &\approx& \mfrac{2}{(\vec{k}^{2}\,
+\, \lambda^{2})^{3/2}}\,[\, 1\,-\,
\mfrac{1}{4}\, \lambda^{2}\,]\nonumber\\
\mfrac{1}{B^{2}\,\sqrt{(1+B)^{2}\,-\,1} } &\approx& \mfrac{4}{(\vec{k}^{2}\,
+\, \lambda^{2})^{5/2}}\,[\, 1\,-\, \mfrac{1}{12}\, \vec{k}^{2}\, -\,
\mfrac{1}{3}\, \lambda^{2}\,]
\end{eqnarray}
The previous expansions are almost all
we need to extract the logarithmic infrared divergence of
our one-loop integrals at every order in the velocity.

As we saw in appendix A, the only infrared divergent integrals are
${\rm Si}^{(21)}(m)$, ${\rm Cs}^{(11)}(m)$ and ${\rm Id}^{(11)}(m)$,
defined in eq.(\ref{defsin}). In order to numerically compute these
integrals, we find their infrared behaviour using (\ref{expanb}) and
then construct the corresponding regularizing integrals which are
\begin{eqnarray} \label{aab}
{\rm Si}^{(21)}(m)_{IR} &=& 4^{m+1}\,
\int_{-\pi}^{\pi}\, d^{3}k\, \mfrac{k_{3}^{2+2m}}
{(\vec{k}^{2}\, +\, \lambda^{2})^{5/2+m}}\, \theta(\pi^{2}-k^{2})\nonumber\\
{\rm Cs}^{(11)}(m)_{IR} &=& 2\, 4^{m}\, \int_{-\pi}^{\pi}\, d^{3}k\,
\mfrac{k_{3}^{2 m}}{(\vec{k}^{2}\, +\, \lambda^{2})^{3/2+m}}\,
\theta(\pi^{2}-k^{2})\nonumber\\
{\rm Id}^{(11)}(m)_{IR} &=& 2\, 4^{m}\, \int_{-\pi}^{\pi}\, d^{3}k\,
\mfrac{k_{3}^{2 m}}{(\vec{k}^{2}\, +\, \lambda^{2})^{3/2+m}}\,
\theta(\pi^{2}-k^{2})
\end{eqnarray}
where we perform the integration on a 3-sphere of radius $\pi$ to take
advantage of spherical symmetry. Of course any other radius $R>0$ would be
equally good.

The three-dimensional integrals in (\ref{aab}) are much simpler to
be evaluated than the original ones. In fact, the best thing we can do is
to separate the radial and angular integrations expanding
$k_{3}$ as a Gegenbauer series
\be \label{aa}
k_{3}^{n}\, =\, k^{n}\, \mfrac{\Gamma(\nu)\, n\!}{2^{n}}\, \sum_{j=0}^{[n/2]}\,
\mfrac{(n-2\, j\, +\, \nu)}{j!\, \Gamma(1+\nu+n-j)}\, C^{\nu}_{n-2
j}(\hat{k}\cdot\hat{e}_{3})
\ee
where $\nu$ is related to the space dimension $D$ by $D=2(\nu+1)$, therefore
$\nu=1/2$ in our case.
Inserting (\ref{aa}) into (\ref{aab}), we have
\begin{eqnarray}
{\rm Si}^{(21)}(m)_{IR} &=& 4^{m+1}\, \mfrac{2}{(2m+3)}\,
I_{R}(m+2,m+2)\nonumber\\
{\rm Cs}^{(11)}(m)_{IR} &=& 4^{m+1}\, \mfrac{1}{(2m+1)}\,
I_{R}(m+1,m+1)\nonumber\\
{\rm Id}^{(11)}(m)_{IR} &=& 4^{m+1}\, \mfrac{1}{(2m+1)}\, I_{R}(m+1,m+1)
\end{eqnarray}
where $I_{R}(\alpha,\beta)$ is the following radial integral
\be
I_{R}(\alpha,\beta)\, =\, \int_{0}^{\pi}\, d k\, \mfrac{k^{2 \alpha}}
{(\vec{k}^{2}\, +\, \lambda^{2})^{3/2+\beta-1}}
\ee
The integration limits are a consequence of the Heviside function introduced
in (\ref{aab}).

The radial integral $I_{R}(m,m)$ can be evaluated by noting that
\be
I_{R}(m,m)\, =\, I_{R}(m-1,m-1)\, +\, \mfrac{\lambda^{2}}{(3/2\, +\, m - 2)}\,
\mfrac{d}{d\, \lambda^{2}}\, I_{R}(m-1,m-1)
\ee
along with
\be
I_{R}(1,1)\, =\, \log(2 \pi)\, -\, 1\, -\, \log(a \lambda)
\ee
The result is
\be
I_{R}(m,m)\, =\, \log(2 \pi)\, -\, \sum_{n=0}^{m-1}\, \mfrac{1}{(2 n +1 )}\,
-\, \log(a \lambda)
\ee
Putting all the formulas together, we arrive at the following infrared
subtracted basic integrals
\begin{eqnarray}      \label{apex1}
{\rm Si}^{(21)}(m) &=& \mfrac{1}{2 \pi}\,
\int_{-\pi}^{\pi}\, d^{3}k\, \left[\,
\mfrac{\sin^{2}(k_{3})}{B^{2}\,
\sqrt{(2+B) B}}\, \Xi^{m}(B)\, -\, 4^{m}\, \mfrac{4\, k_{3}^{2(m+1)}
\, \theta(\pi^{2}-k^{2})}{
\left( k^{2}\, +\, \lambda^{2}\right)^{5/2+m}}\, \right]\nonumber\\
&+& \mfrac{2\, 4^{m+1}}{(2 m + 3)}\, \left[ \log(2 \pi)\, -\,
\sum_{n=0}^{m+1}\, \mfrac{1}{(2 n +1 )} \,-\, \log(a \lambda) \right]\\
\label{apex11}
{\rm Cs}^{(11)}(m) &=& \mfrac{1}{2 \pi}\,
\int_{-\pi}^{\pi}\, d^{3}k\, \left[\, \mfrac{\cos^{2}(k_{3}/2)}{B\,
\sqrt{(2+B) B}}\, \Xi^{m}(B)\, -\, 4^{m}\, \mfrac{2\, k_{3}^{2m}
\, \theta(\pi^{2}-k^{2})}{
\left( k^{2}\, +\, \lambda^{2}\right)^{3/2+m}}\, \right]\nonumber\\
&+& \mfrac{4^{m+1}}{(2 m +1)}\, \left[ \log(2 \pi)\, -\,
\sum_{n=0}^{m}\, \mfrac{1}{(2 n +1 )} \,-\, \log(a \lambda ) \right]\\
\label{apex12}
{\rm Id}^{(11)}(m) &=& \mfrac{1}{2 \pi}\,
\int_{-\pi}^{\pi}\, d^{3}k\, \left[\, \mfrac{1}
{B\, \sqrt{(2+B) B}}\, \Xi^{m}(B)\, -\, 4^{m}\, \mfrac{2\, k_{3}^{2m}\,
\theta(\pi^{2}-k^{2})}{
\left( k^{2}\, +\, \lambda^{2}\right)^{3/2+m}}\, \right]\nonumber\\
&+& \mfrac{4^{m+1}}{(2 m +1)}\, \left[ \log(2 \pi)\, -\,
\sum_{n=0}^{m}\, \mfrac{1}{(2 n +1 )} \,-\, \log(\lambda a) \right]
\end{eqnarray}
Using the previous equations, we have calculated the numerical values of
these divergent integrals which are tabulated in Table A.2. Moreover,
the coefficients of
the logarithmic divergence determine the anomalous dimension of the heavy-heavy
quark current and the running of the derivatives of the Isgur-Wise
function.

\end{document}